\documentclass[useAMS,usenatbib]{mn2e}
\usepackage{hyperref}
\usepackage{amssymb,amsmath,esint}
\usepackage{graphicx}
\usepackage{epstopdf}
\usepackage{threeparttable}
\usepackage{multirow}
\usepackage{mathptmx}
\usepackage{color}
\newcommand{\bea}{\begin{eqnarray}}
\newcommand{\eea}{\end{eqnarray}}
\newcommand{\beq}{\begin{equation}}
\newcommand{\eeq}{\end{equation}}

\newcommand{\Lam}{\ensuremath{\Lambda}}

\title[Jubilee ISW II: superstructure imprints]{The Jubilee ISW Project II: observed and simulated imprints of voids and superclusters on the cosmic microwave background}
\author[S. Hotchkiss et al.]{S.~Hotchkiss$^{1}$,\thanks{s.a.hotchkiss@sussex.ac.uk} S.~Nadathur$^{2}$,\thanks{seshadri.nadathur@helsinki.fi}S.~Gottl\"ober$^3$, I.~T.~Iliev$^1$, A.~Knebe$^4$, \newauthor W.~A.~Watson$^1$, G.~Yepes$^4$
\\
$^1$Department of Physics and Astronomy, University of Sussex, Brighton, BN1 9QH, UK \\
$^2$Department of Physics, University of Helsinki and Helsinki Institute of Physics, P.O. Box 64, FIN-00014, University of Helsinki, Finland\\
$^3$ Leibniz-Institute for Astrophysics, An der Sternwarte 16, 14482
Potsdam, Germany\\
$^4$ Departamento de F\'isica Te\'orica, Modulo C-XI, Facultad de Ciencias, 
Universidad Aut\'onoma de Madrid, 28049 Cantoblanco, Madrid, Spain\\
}

\begin{document}

\date{\today}

\pagerange{\pageref{firstpage}--\pageref{lastpage}}

\label{firstpage}

\maketitle

\begin{abstract}
We examine the integrated Sachs-Wolfe (ISW) imprint of voids and superclusters on the cosmic microwave background. We first study results from the Jubilee $N$-body simulation. From Jubilee, we obtain the full-sky ISW signal from structures out to redshift $z=1.4$ and a mock luminous red galaxy (LRG) catalogue. We confirm that the expected signal in the concordance \Lam CDM model is very small and likely to always be much smaller than the anisotropies arising at the last scattering surface. Any current detections of such an imprint must, therefore, predominantly arise from something other than an ISW effect in a \Lam CDM universe. Using the simulation as a guide, we then look for the signal using a catalogue of voids and superclusters from the Sloan Digital Sky Survey. We find a result that is consistent with the \Lam CDM model, i.e. a signal consistent with zero.
\end{abstract}


\maketitle

%--------------Introduction------------------%

\section{Introduction}
\label{sec:Intro}

An impressive array of observational evidence \citep[e.g.,][]{Percival:2009xn,Beutler:2011hx,Blake:2011en,Anderson:2012sa,Hinshaw:2012aka,Hou:2012xq, Sievers:2013ica,Planck:CosmoParams} has led to the \Lam~Cold Dark Matter (\Lam CDM) model of cosmology being viewed as the standard cosmological model. One consequence of this \Lam CDM model is that, when the cosmological constant begins to affect the universe, gravitational potentials in the universe decay. This decay leaves an imprint on the cosmic microwave background (CMB) in the form of a pattern of secondary anisotropies formed when CMB photons traverse regions of over- or underdensity. This effect is known as the integrated Sachs Wolfe (ISW) effect.

In an Einstein-de Sitter universe containing only pressureless dust gravitational potentials do not decay at linear order, so the ISW effect is too small to detect amidst the primary anisotropies in the CMB produced at the last scattering surface. The nature of the observed ISW signal and its dependence on density and redshift are therefore good probes of how our Universe might deviate from this homogeneous pressureless state, whether through a cosmological constant or some new physics \citep{Crittenden:1995ak}.

Unfortunately, even in \Lam CDM, the ISW effect is also too small to be observed directly. It could therefore only ever be observed through the correlation of the distribution of some tracer of the density field in the local Universe and the CMB. Assuming Gaussian initial conditions and linear growth of density fluctuations, the optimal detection method is a full cross-correlation of maps of the density field and temperature anisotropies \citep{Afshordi:2004kz}. There is now a large collection of works using this method to attempt to measure the effect, with some reasonably significant detections claimed through the use of combinations of tracers \citep{Giannantonio:2008zi,Ho:2008bz,Giannantonio:2012aa}; although, see \citet{Dupe:2010zs}. Future surveys will be able to obtain greater sensitivity and will be able to constrain the evolution of dark energy, through e.g. the equation of state \citep{Douspis:2008xv}.

However, if the real universe is not well described by the simplest \Lam CDM model then a cross-correlation may not be the optimal method to detect an ISW-like effect. Therefore, it is worth considering other detection strategies. One such method was employed by \citet*{Granett:2008ju} (hereafter G08), who stacked cutouts of the CMB along the lines of sight of superstructures (large-scale voids and superclusters) identified in a catalogue of LRGs in the Sloan Digital Sky Survey (SDSS). By analysing these stacked images they found a highly significant ($4.4\sigma$) temperature signal at the locations of superstructures. This is a much higher significance detection than any conventional cross-correlation study has been able to obtain from a single data set, including those that have also used LRGs \citep{Padmanabhan:2004fy,Cabre:2006qm,Granett:2008dz,Ho:2008bz,Giannantonio:2012aa}. Whereas the detection was initially claimed as a verification of the \Lam CDM model, the measured value has subsequently been shown to be significantly larger than \Lam CDM would predict \citep*{Hunt:2008wp,Nadathur:2011iu,Flender:2012wu,HernandezMonteagudo:2012ms}. This potentially calls aspects of \Lam CDM into question, at least at the redshifts, densities and distance scales probed by this measurement.

Given the importance of such a result it is important to examine it from as many avenues as possible. Firstly, although it is clear that the measured signal is too big for \Lam CDM it is still unclear what the \emph{precise} \Lam CDM expectation is. This is because previous analyses have all only been able to derive upper bounds on the maximum possible signal. If this observation is to be explained by some sort of new physics then a precise understanding of the expected \Lam CDM behaviour will be required for quantitative comparisons. Equally, it may be that the \Lam CDM signal can be shown to be eventually measurable---with more sky coverage and perhaps a different tracer population---which could provide a different observational opportunity.

In this work we analyse the ISW temperature shift due to voids and superclusters in a mock sample of LRGs produced from the Jubilee simulation \citep{Watson:2013mea}. An earlier study has also examined the expected signal from voids alone in a \Lam CDM $N$-body simulation \citep[][hereafter C13]{Cai:2013ik}, but Jubilee is the first simulation which simultaneously provides a small enough halo mass resolution to model the LRG population in the simulation as well as a large enough volume to encompass the entire redshift range where the ISW effect is relevant. In fact, Jubilee is large enough to cover the entire sky out to $z=1.4$ without repetition of the simulation box and also contains the largest modes that are often neglected in smaller simulations but still contribute very significantly to the largest scales. This allows for the first time a complete modelling of the observation pipeline. We study the optimal parameters for detection, but find that the expected \Lam CDM signal is always too small to be detected above the primary CMB anisotropies, thus confirming the previous analytical results.

Having calibrated our expectations using the simulation, we then turn to catalogues of both voids and superclusters presented by \citet{Nadathur:2014a} (hereafter NH14) which were identified in the real spectroscopically selected sample of SDSS LRGs the Jubilee mock catalogue is intended to mimic. We find no evidence for an imprint on the CMB that could have arisen from either voids or superclusters. This null result is consistent with \Lam CDM.

Our result differs a little from the earlier reported claim of a $\sim 2\sigma$ significant detection of a temperature decrement \emph{from voids alone} (C13), which corresponds to an amplitude of signal that is several times larger than the \Lam CDM expectation. This study used the same tracer galaxy sample as we do, but a somewhat different catalogue of voids. We are able to reproduce this result, but argue that the signal does not exhibit a significant trend with void properties. This is consistent with the explanation of the detection as having arisen through random noise rather than a real physical effect. We also observe that the tentative detection in C13 does not match the behaviour of that reported in G08. In fact, applying the same superstructure selection criteria as in G08 to either our catalogues or the catalogue in C13 produces a null result consistent with \Lam CDM.

The original high significance detection made by G08, made using a sample of photometric LRGs at a higher redshift ($0.4<z<0.75$) still stands \citep{Planck:ISW}, and remains unexplained. However, no similar signal is seen in LRGs at a lower redshift, where the ISW effect is expected to be larger.

The outline of the rest of this paper is as follows. In section \ref{sec:JUB} we describe the properties of the Jubilee simulation and the simulated ISW maps used in our modelling. In section \ref{sec:Structs} we describe the methods we use to produce superstructure catalogues from the simulation, and in section \ref{sec:ISWsim} we examine the expected ISW imprint from these structures and the optimal detection strategy. In section \ref{sec:ISWdata} we use the same strategy to look for the stacked ISW signal in SDSS data, before comparing our findings with previous work in section \ref{sec:prevwork} and summarising our results in section \ref{sec:conclude}.

%-------------Jubilee Project------------%

\section{The Jubilee ISW project}
\label{sec:JUB}
To assess the \Lam CDM \emph{expectation} for the stacked ISW signal, we analyse data from the Jubilee ISW project \citep{Watson:2013cxa}. The Jubilee ISW project is built upon the Jubilee simulation, a \Lam CDM $N$-body simulation of large-scale structure presented in \citet{Watson:2013mea}. It consists of $6000^3$ particles in a volume of $(6 h^{-1} \mathrm{Gpc})^3$, corresponding to individual particle masses of $7.49 \times 10^{10} \rm{M}_\odot$ and a minimum resolved halo mass (with $\simeq 20$ particles) of $\simeq 1.5 \times 10^{12} h^{-1} M_\odot$. The initial conditions of the simulation were set at redshift $z=100$ and used the following cosmological parameters, motivated by the 5-year WMAP results \citep{Dunkley:2008ie}: $\Omega_\rmn{m}=0.27$, $\Omega_\Lam=0.73$, $h=0.7$, $\Omega_\rmn{b}=0.044$, $\sigma_8=0.8$ and $n_\rmn{s}=0.96$.

Its large size and relatively high resolution means the Jubilee simulation is ideal for analysing the ISW effect. Specifically, the large box size allows a light cone to be constructed that requires no tiling of the simulation box out to a redshift of $z=1.4$. Therefore, full sky maps of the temperature anisotropies induced by the ISW effect can be constructed that will not suffer from a cutoff of power on the largest angular scales. The halo mass resolution allows halo occupation distribution (HOD) modelling of tracers such as LRGs. This allows for a complete modelling of the observation pipeline in stacking analyses. Although other smaller ISW simulations exist with similar minimum resolvable halo mass \citep[][C13]{Cai:2010hx}, they are unable to match the volume of Jubilee (see Table \ref{tab:thetable} for a comparison of properties). 

\begin{table*}
\begin{minipage}{166mm}
\caption{Properties of simulations used in analyses of stacked ISW signal.}
\begin{threeparttable}
%\begin{centering}
\label{tab:thetable}
\begin{tabular}{@{}lrcccc}
\hline
ISW Simulation & & Box size (${\rm Mpc}/h$) & Particle mass ($M_\odot/h$) & Minimum halo mass ($M_\odot/h$) &\\
\hline
Jubilee (this paper) & &$6000$ & $7.49 \times 10^{10}$ &  $1.49 \times 10^{12}$&
\\
 &\multirow{3}{*}{\Huge \{ }& $1500$ & $2.09 \times 10^{11}$ & $4.18 \times 10^{12}$&\multirow{3}{*}{\Huge \} }\\
\cite{Cai:2013ik} (C13)& &$1000$ & $6.20 \times 10^{10}$ & $1.24 \times 10^{12}$&\\
& &$250$ & $7.7 \times 10^{9}$ & $1.54 \times 10^{11}$&\\ 
\citet{HernandezMonteagudo:2012ms} & &$1500$  & $5.56 \times 10^{11}$  & $1.11 \times 10^{13}$&\\
\citet{Flender:2012wu}\tnote{1} & &$1000$ & $6.78 \times 10^{9}$ & $1.35 \times 10^{11}$&\\
\hline
\end{tabular}
\begin{tablenotes}
\item [1] Used ISW maps produced from simulation in \cite{Cai:2010hx}.
\end{tablenotes}
\label{table:voidvolfrac}
%\end{centering}
\end{threeparttable}
\end{minipage}
\end{table*}

\subsection{The ISW effect in Jubilee}
\label{sec:ISWJub}

The Jubilee maps of the ISW-induced temperature anisotropies were constructed using a semi-linear approach introduced by \citet{Cai:2010hx}. In that work it was demonstrated that this approximation is very accurate up to $l\lesssim 50$. At smaller scales and at very high redshifts the non-linear Rees-Sciama effect begins to dominate over the linear ISW effect. However, both the Rees-Sciama and ISW effect at these small scales and early times are overwhelmed by the large scale, late time, linear ISW effect we are interested in.

The ISW-induced temperature fluctuations in the CMB are given by
\bea
  \label{eq:ISW}
  \Delta T(\hat{n}) = 2\overline{T} \int \dot{\Phi}(r,\hat{n})\, a\, dr\,,
\eea
where $\dot{\Phi}$ is the derivative of the gravitational potential with respect to time, $\overline{T}$ is the mean temperature of the CMB, and the integral is along the photon path from the last scattering surface.

To extract a prediction for $\dot{\Phi}$ we make use of its relationship to $\delta$, through the Poisson equation:
\bea
  \label{eq:Poiss}
  \Phi_k=-\frac{3}{2}\Omega_\rmn{m} \frac{H_0^2}{k^2}\frac{\delta_k}{a}\,,
\eea
which gives, for the time derivative of $\Phi$,
\bea
  \label{eq:Poissdot}
\dot{\Phi}_k = \frac{3}{2}\Omega_\rmn{m} \frac{H_0^2}{k^2}\left[\frac{H}{a}\delta - \frac{\dot{\delta}}{a}\right].
\eea
The semi-linear approach introduced by \citet{Cai:2010hx} takes the full non-linear density field $\delta$ in the formulae above, but assumes that the time derivative $\dot{\delta}$ is given by linear theory, i.e. $\dot{\delta}(t) = \dot{D}(t)\delta_0$, where $D(t)$ is the linear growth function. It is then possible to use equations \eqref{eq:Poiss} and \eqref{eq:Poissdot} to obtain
\bea
  \label{eq:LAV}
  \dot{\Phi}= -\Phi\, H(t) \left[1-\beta(t)\right],
\eea
where $\beta=d\ln{D}/d\ln{a}$. During matter domination $\beta =1$. It becomes $<1$ when dark energy begins to noticeably affect the universe. We obtain $\Phi$ from simulation outputs at 20 separate redshifts between $z=0$ and $z=1.4$ by solving the Poisson equation in Fourier space. Even at the smallest scales that are resolved by the simulation and at the fully non-linear level, the gravitational potential changes slowly. Therefore, at intermediate redshifts, $\Phi$ is obtained at each point in space using a linear interpolation between the values at the same location in the two nearest redshift outputs. We propagate light rays through the simulation box and use equation \eqref{eq:ISW} to obtain the sky maps of the temperature shift along different directions as seen by a centrally located observer. These maps are pixellised using the HEALPix package \citep{Gorski:2004by}\footnote{\url{http://healpix.jpl.nasa.gov}} at resolution $N_\rmn{side}=512$. We also produce ISW sky maps due to structures in the individual redshift shells, allowing us to determine not just the total ISW signal but also the contributing redshift interval for our set of superstructures.

Further details of the method can be found in \citet{Watson:2013cxa}.

\subsection{LRG modelling}
\label{sec:LRGs}

A full modelling of the stacking analysis with Jubilee requires realistic mock galaxy catalogues similar to those in which real voids and superclusters are identified. We make use of the mock LRG catalogues introduced by \citet{Watson:2013cxa}. These are constructed by using an HOD model based on the results of \citet{Zheng:2008np} to populate haloes in our simulation. This HOD model is itself calibrated on a sample of SDSS LRGs with $g$-band magnitudes $M_g<-21.2$ between redshifts of $0.16$ and $0.44$ \citep{Eisenstein:2005su}. LRGs typically reside in haloes of mass in excess of $\sim10^{13}\;h^{-1}\rmn{M}_\odot$ \citep{Zheng:2008np,Wen:2012tm,Zitrin:2012}, which is well above the resolution limit of Jubilee. We take model parameters from \citet{Zheng:2008np} assuming that LRGs are the brightest cluster galaxy (BCG) in their respective halo, thus neglecting the small fraction ($\sim5\%$) of LRGs that are satellite galaxies.

We assign luminosities to the LRG population based on the host halo masses \citep{Zheng:2008np}, after accounting for the varying steepness of the mass-luminosity relationship as a function of halo mass \citep[see][for details]{Watson:2013cxa}. We apply a log-normally distributed random scatter between the LRG location and that of the dark matter density peak as seen for BCGs in the results of \citet{Zitrin:2012}. We assume the LRG peculiar velocity to be the same as that of its host halo and include this as a Doppler correction term to the `observed' redshift of the LRGs. We then convert these `observed' redshifts into `observed' LRG positions in comoving coordinates using our fiducial cosmology. Further details and discussion of all of these modelling steps are provided in \citet{Watson:2013cxa} where the LRG catalogues were introduced.

We apply magnitude and redshift cuts to this sample to construct two mock full-sky LRG samples designed to match the properties of the actual (quasi-) volume-limited SDSS DR7 LRG samples presented in \citet{Kazin:2010}. We select mock LRGs with $-23.2<M_g<-21.2$ and redshift $0.16<z<0.36$ to create the `Jubilee Dim' (JDim) sample, and those with $-23.2<M_g<-21.8$ and $0.16<z<0.44$ to create the `Jubilee Bright' (JBright) sample. These properties are intended to match the Dim and Bright subsamples of \citet{Kazin:2010}, from which the catalogues of voids and superclusters presented by NH14 were drawn. Beyond the scope of this work alone, these mock catalogues are useful for comparing properties \citep[e.g., radial density profiles][]{Nadathurforth} of these superstructures in simulation and SDSS.

Unfortunately, some of the superstructures listed in NH14 were identified in populations of tracer galaxies less massive than LRGs, which cannot be resolved fully by Jubilee. We therefore restrict ourselves primarily to examining and modelling superstructures generated from the two LRG catalogues used by NH14 alone.

%-------------Catalogues------------%

\section{Identifying voids and superclusters}
\label{sec:Structs}

\subsection{Structures in Jubilee}
\label{sec:Jubstruc}

We identify `superstructures' (large voids and overdensities, or `superclusters') in the JDim and JBright mock LRG samples using a modified version of the {\small ZOBOV} void-finding algorithm \citep{Neyrinck:2007gy}, according to the prescription laid out in NH14, which for completeness we briefly recap here. 

{\small ZOBOV} first estimates the local galaxy density by means of a three-dimensional Voronoi tessellation of the galaxy distribution. The Voronoi cell of each galaxy consists of the region of space closer to it than to any other galaxy, and a density value equal to the inverse of the Voronoi cell volume is assigned to each location. As the tessellation step is designed to operate on a cubic box whereas our mock LRGs are distributed along a light cone and are therefore contained within a spherical shell around the central observer, we follow the usual procedure of enclosing the mock galaxies within a buffer of boundary particles at both the lower and upper redshift caps. It is important for a robust density reconstruction that these boundary particles are sufficiently densely packed to ensure that no Voronoi cells leak outside the survey volume and that `edge' LRGs adjacent to boundary particles are identified and appropriately handled; our method here exactly follows that outlined in detail in NH14. 

Within this reconstructed density field, {\small ZOBOV} identifies local minima and their associated catchment `zones', using the watershed transform. For any individual Voronoi cell this is done by first finding the set of cells it is adjacent to; then, the initial cell is put into the same zone as the smallest density adjacent cell. This is done iteratively, until the smallest density adjacent cell is a density minimum, which then defines the ``core'' of the zone.

The void catalogues are then produced as follows from the full set of {\small ZOBOV} catchment zones. Firstly, a seed zone is chosen, starting with the minimum density zone in the survey. Then, neighbouring sets of zones are added to the seed zone to form a void, in order of their ``linking density'' $\rho_\rmn{link}$ to the void being formed. The linking density between two distinct sets of zones is defined as the minimum density of the cells on the border between the sets. Note that more than one zone can be added at one link density if multiple zones are linked to \emph{each other} by a smaller linking density. This process is then continued until specific stopping conditions are satisfied, at which point the growth of the void is stopped. Then, the zone with the next smallest density is found, and if it is not already part of a void, it forms the seed for a new void.  The ``stopping conditions'' chosen depend on the desired properties of the output voids.

Type1 voids are defined according to the following criteria: (1) to qualify as a starting seed, a zone must have a minimum density $\rho\rmn{min}<0.3\overline{\rho}$, where $\overline{\rho}$ is the mean density; (2) zone merging is halted once either (a) $\rho_\rmn{link}$ exceeds $\overline{\rho}$, or (b) the \emph{density ratio} $r$ between the link density and the minimum density of the zone(s) being added exceeds a threshold value $r=2$ above which that zone is considered an independent void. As discussed in NH14, fewer than $0.7\%$ of the spurious `voids' identified in random point distributions match these characteristics, so Type1 voids are statistically distinct from such a noise-induced population at the $3\sigma$ equivalent confidence level.

Type2 voids are defined even more conservatively, requiring $\rho_\rmn{min}<0.2\overline{\rho}$ to qualify as a void, as well as restricting merging to the case $\rho_\rmn{link}<0.2\overline{\rho}$.

{\small ZOBOV} can also be applied to the inverse of the density field (i.e., directly to the Voronoi volumes) to obtain a set of over-densities, or `superclusters'. This process is similar to the VOBOZ algorithm \citep{Neyrinck:2004gj}, but without VOBOZ's final step, which attempts to use particle velocity information to return only virialised, spherically symmetric haloes. To define our set of superclusters, we again follow the criteria laid out in NH14: only zones with $\rho_\rmn{max}>22\overline{\rho}$ are allowed to start forming superclusters, and neighbouring zones are added until a set of zones is encountered with $\rho_\rmn{link}<\overline{\rho}$ or $r>16.3$ (with $r$ in this case defined as the ratio of the maximum to link density). These criteria are applied to ensure that the population of superclusters are distinct from Poisson point distributions. They are the equivalent to the conditions defining Type1 voids.

Two additional points are worth noting. Firstly, {\small ZOBOV} actually normalizes all densities in units of the mean. Rather than use the overall mean for the full sample, we correct for the redshift-dependent variation of the mean density, determined in broad redshift bins. This is because the real LRG distribution has a radial dependence in the mean number density due to the survey selection function; as a result our simulated LRG samples have a radial dependence too. Secondly, we do not attempt to mimic the SDSS survey mask with Jubilee, instead treating it as a full-sky galaxy catalogue. Clearly, applying the survey mask would most closely match the actual observation; however, it also reduces the number of structures available for analysis, thus limiting the statistical power available in Jubilee.\footnote{This would be true even if multiple SDSS-like windows were modelled over the sky because of the conservative treatment of survey edges.} Whereas it is certainly interesting to examine how the statistical properties of structures depend on a survey window's precise shape and size, we are interested here in the ISW signal along a superstructure's line of sight, \emph{given that superstructure's properties}. This relationship will be much less sensitive to the properties of the observed window because the CMB photons that experience the ISW effect are not affected by survey masks. This will be especially true for $\rho_{\rm min}$ and $\rho_{\rm max}$. However, because we scan over the filter rescaling weight in our results, any mask-induced error in determining $R_{\rm eff}$ will also be negated. Not applying a mask also allows us to make predictions for the signal that may be observed using future surveys with greater sky coverage.

In total we find $657$ Type1 voids, $377$ Type2 voids and $1104$ superclusters in JDim, and $342$ Type1 voids, $166$ Type2 voids and $774$ superclusters in JBright. For each structure we record the sky position of its volume-weighted barycentre, its effective radius $R_\rmn{eff}$---defined as the radius of a sphere with the same volume as the structure---and the angle subtended on the sky by a sphere of this radius located at the radial distance of the barycentre, $\Theta_\rmn{eff}$. 

The {\small ZOBOV}-based void-finding method outlined above differs from the one previously used in \cite{Watson:2013mea} to analyse extreme events in Jubilee, where voids were defined as the spherical regions in the simulation that do not contain any halo above a specified threshold mass. This definition has the advantage that it is easy to model theoretically: the halo mass function gives the abundances of haloes of a given mass and the galaxy-galaxy correlation function dictates how they cluster. From this, the probability that a spherical region contains no haloes above a mass threshold is straightforwardly obtained. However, it has the disadvantage that locations of such voids will only be loosely correlated with the true \emph{large-scale} underdensities in the total matter field, because such regions may still contain small numbers of isolated haloes with large masses. Similarly, large-scale over-densities will often not contain any of the \emph{most} massive haloes. It is these large-scale density fluctuations that we are interesed in, because they will have the largest ISW imprints and {\small ZOBOV} is better at locating them. Such regions are also easier to define, using {\small ZOBOV}, in real galaxy data. However, we note that these advantages come at the cost of making the superstructure populations found by {\small ZOBOV} more difficult to model theoretically.

\subsection{Structures in SDSS data}
\label{sec:SDSSstructs}

In this work for observational comparison with simulation results we use the catalogues of voids and superclusters identified in SDSS DR7 galaxy samples presented in NH14. The work of NH14 borrowed motivation and some methodology from an earlier attempt to build a catalogue of voids from the same data \citep{Sutter:2012wh}. However, as explained in NH14 and \citet{NH:2013b}, the \citet{Sutter:2012wh} void catalogue, although worthwhile in motivation, suffered from a number of problems, including boundary contamination and the mistaken classification of \emph{overdense} structures as `voids', requiring the construction of a new catalogue. 

For the most part, we will restrict ourselves to the voids and superclusters identified in the two LRG galaxy samples referred to by NH14 as \emph{lrgdim} and \emph{lrgbright}. These are extensions of the Dim and Bright samples of \cite{Kazin:2010} which include additional galaxies from the southern Galactic sky. They therefore have exactly the same magnitude and redshift cuts: $-23.2<M_g<-21.2$ and $0.16<z<0.36$ for \emph{lrgdim},  $-23.2<M_g<-21.8$ and $0.16<z<0.44$ for \emph{lrgbright}---as those used to create the mock samples JDim and JBright. Modelling uncertainties, in particular the fact that the HOD model does not allow for redshift evolution of the best fit parameters, may introduce small differences in the redshift distribution of the real and mock LRGs. However, since we correct for the redshift-dependence of the local mean density as described in section \ref{sec:Jubstruc}, this difference will not be important. In total there are 70 Type1, 19 Type2 voids and 196 superclusters in \emph{lrgdim}, and 13 Type1 voids, 1 Type2 void and 39 superclusters in \emph{lrgbright}.

In addition, we will also briefly consider superstructures identified by NH14 in the four main galaxy samples drawn from the New York University Value-Added Galaxy Catalog  \citep[NYU-VAGC;][]{Blanton:2004aa} referred to as \emph{dim1}, \emph{dim2}, \emph{bright1} and \emph{bright2}. These have $r$-band magnitudes $M_r < -18.9$, $M_r<-21.35$, $M_r<-21.35$ and  $M_r<-22.05$, and maximum redshift extents $z<0.05$, $z<0.1$, $z<0.15$ and $z<0.2$ respectively. The halo mass resolution of Jubilee is not sufficient to accurately model these populations, so in principle the conclusions we draw about the optimal stacking strategy for structures in Jubilee may not be applicable to these structures. However we note that other results \citep[NH14;][]{Nadathurforth} hint at a universality of void properties independent of the tracer population.

%-------------Next Section------------------%

\section{The stacked ISW signal in simulations}
\label{sec:ISWsim}

We now examine the stacked ISW signal of the superstructures found in the Jubilee simulation. To do this we first extract patches from the simulated ISW maps along the lines of sight of identified superstructures. We then filter each patch using a compensated top-hat filter defined by
\bea
\label{eq:filtdef}
\Delta T(\theta_R) = \frac{\iint_{0}^{\theta_R} T(\theta) d\theta d\phi-\iint_{\theta_R}^{\theta_R^\ast} T(\theta) d\theta d\phi}{\iint_{0}^{\theta_R} d\theta d\phi}\, ,
\eea
and average the resulting filtered temperatures from all patches. Here $\theta$ is the azimuthal angle from the line of sight passing through the centre of each structure, $\theta_R$ is the filter angle and $\theta_R^\ast= \arccos(2\cos(\theta_R)-1)$. This subtracts the average temperature over an annular ring from that within an enclosed disk of equal area,\footnote{For small filter sizes $\theta_R$, the approximation $\theta_R^\ast\simeq\sqrt{2}\theta_R$ suffices.} and is introduced to remove the effect of temperature fluctuations on angular scales larger or smaller than $\theta_R$.

The first use of such a filter in stacked ISW analyses was by G08, who chose a single filter radius $\theta_R$ for all superstructures in their catalogue and then examined the behaviour of the average signal $\overline{\Delta T}$ with this radius, finding a maximum at $\theta_R\sim4^\circ$. However, the original G08 catalogues were limited to 50 voids and superclusters each, with a relatively small distribution of sizes. Given the large number of structures in our catalogues and their very large range of angular sizes, we follow the alternative procedure \citep[][C13]{Ilic:2013cn,Planck:ISW} of rescaling the filter radius for each superstructure in proportion to the angle it subtends on the sky:
\bea
  \label{eq:rescdef}
  \theta_R=\alpha \Theta_\rmn{eff},
\eea
and determining the optimal rescaling ratio $\alpha$ from simulation. This procedure should give a larger expected signal.\footnote{Interestingly, however, when the G08 data is re-analyzed using rescaled filter radii for each void the significance of the measured signal \emph{decreases} \citep{Ilic:2013cn}. This happens because the signal decreases \emph{and} the noise increases.}

Our goal in the following sections is to use Jubilee to examine the expected signal in \Lam CDM and to choose, as far as possible, optimal search parameters to maximise the signal. We then apply these parameters to the real data from Planck and SDSS in the hope of detecting a signal. Although the \Lam CDM expectation in fact turns out to always be too small to be detectable, this procedure importantly avoids the risk of \emph{a posteriori} bias in interpretation of the data. In practice it means our detection strategy is only sensitive to new physical effects which amplify the ISW signal from such superstructures but leave its other characteristics unchanged.

\subsection{Optimizing the filter radius}
\label{sec:filtsize}

\begin{figure}
 \centering   
 \includegraphics[width=0.43\textwidth]{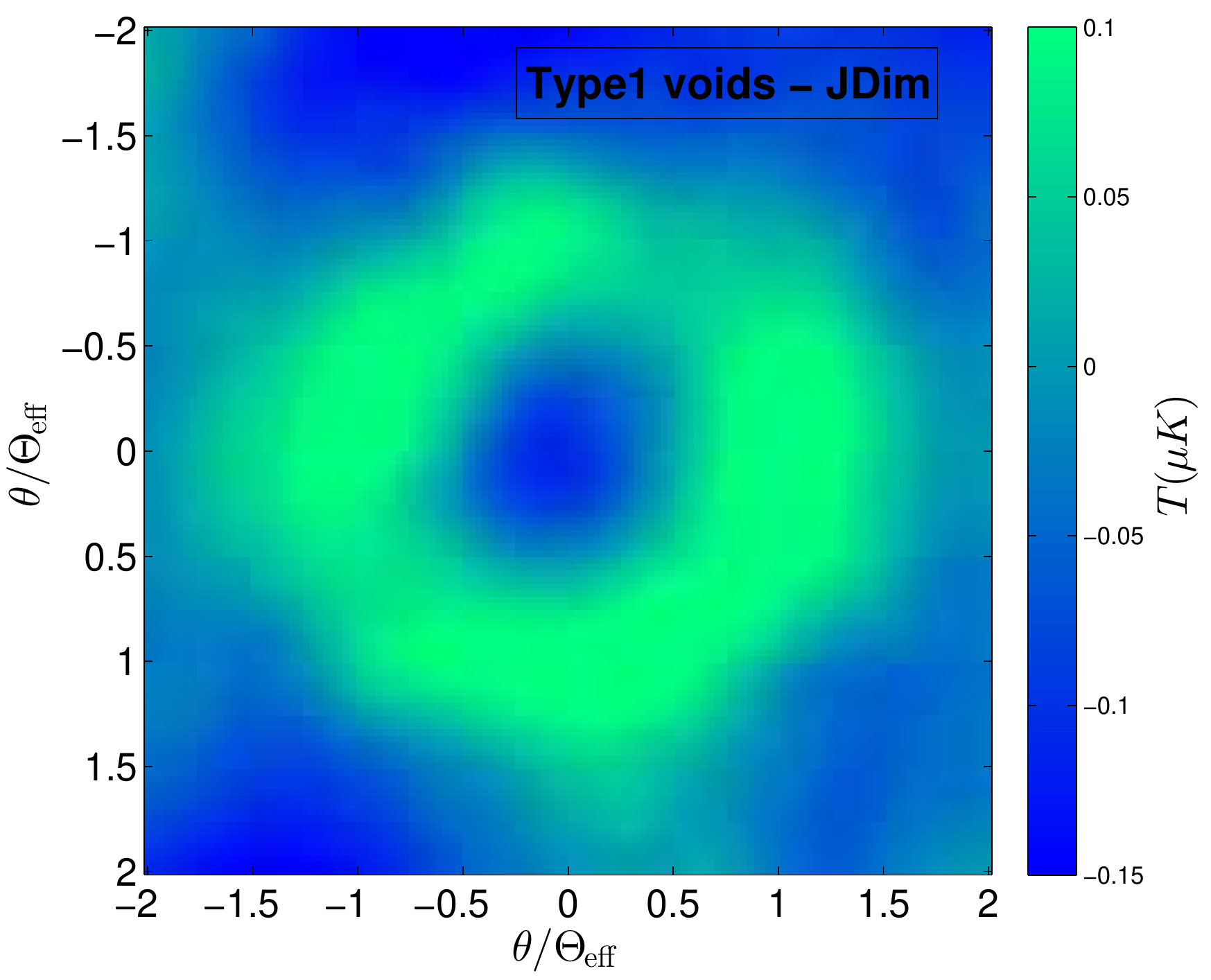}
 \includegraphics[width=0.43\textwidth]{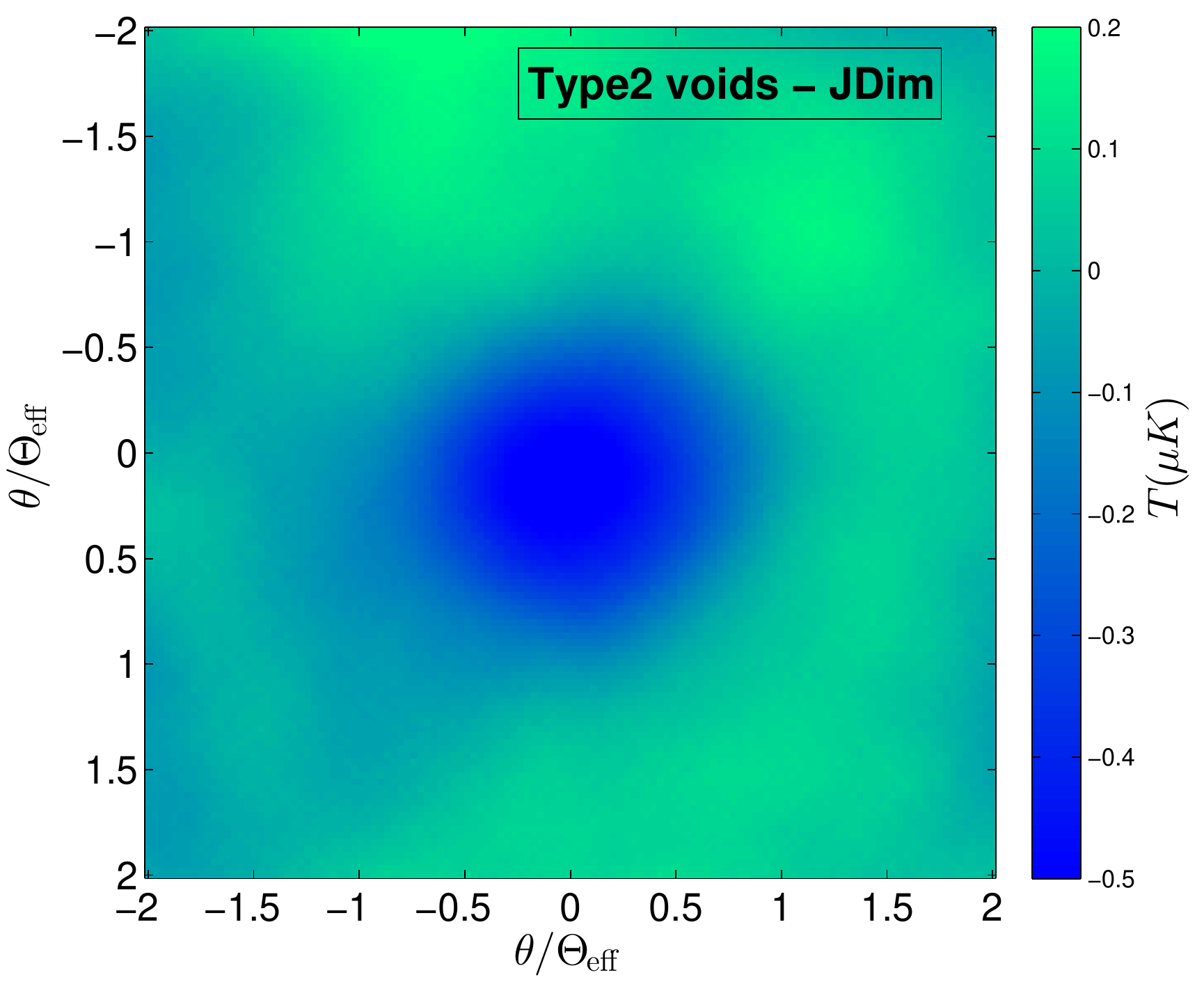}
 \includegraphics[width=0.43\textwidth]{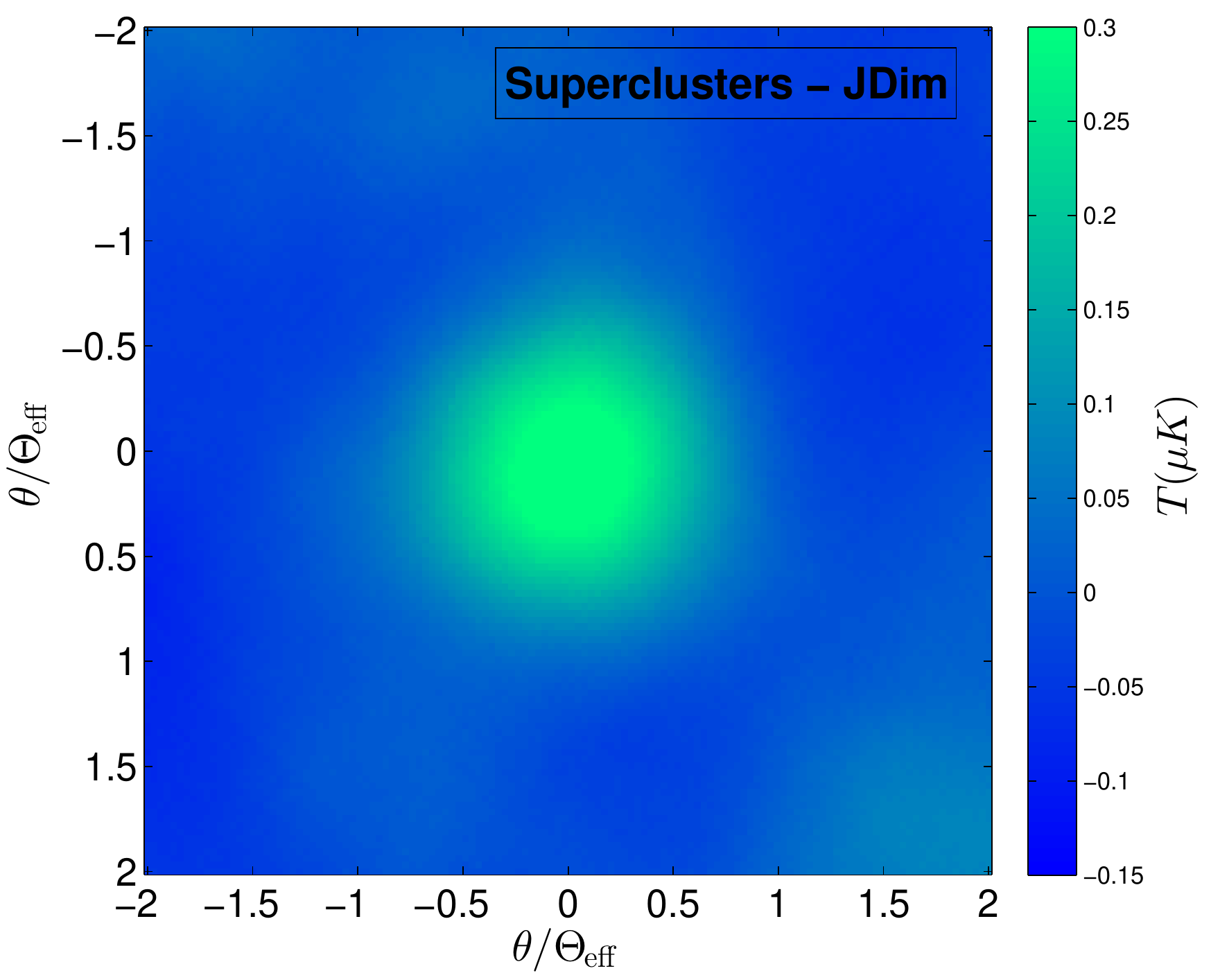}
 \caption{Stacked images of the \emph{unfiltered} ISW temperature anisotropy along the lines of sight of JDim Type1 voids (top), Type2 voids (middle) and superclusters (bottom) found within the Jubilee simulation. Each structure's line of sight patch has been resized proportionately to the effective sky angle subtended by the structure.}
 \label{fig:stackimg}
\end{figure}

In Figure \ref{fig:stackimg} we show the stacked images of the ISW temperature anisotropy along the lines of sight for all Type1 voids, Type2 voids and superclusters in the JDim mock LRG sample. The map patches have been rescaled according to the structure size $\Theta_\rmn{eff}$ before stacking, and we have removed large-scale contributions from multipoles $l\leq10$. Cold and hot spots are visible for the voids and superclusters respectively. Type2 voids, being defined as deeper density minima, show a cold spot that is a factor of two colder than that for Type1. They also lack the characteristic hot ring seen in the top panel. 

\begin{figure}
  \centering
 \includegraphics[width=0.45\textwidth]{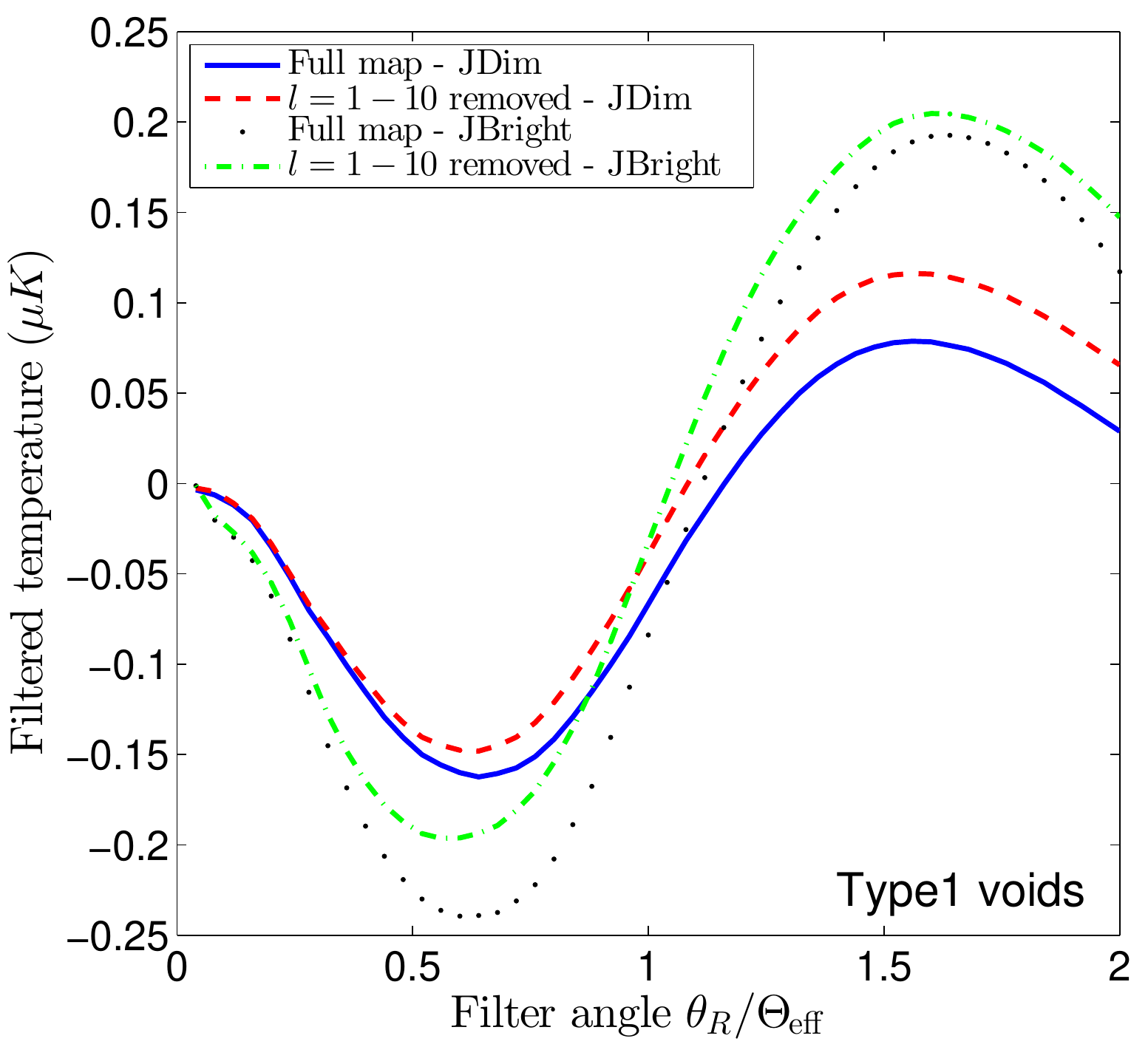}
 \includegraphics[width=0.45\textwidth]{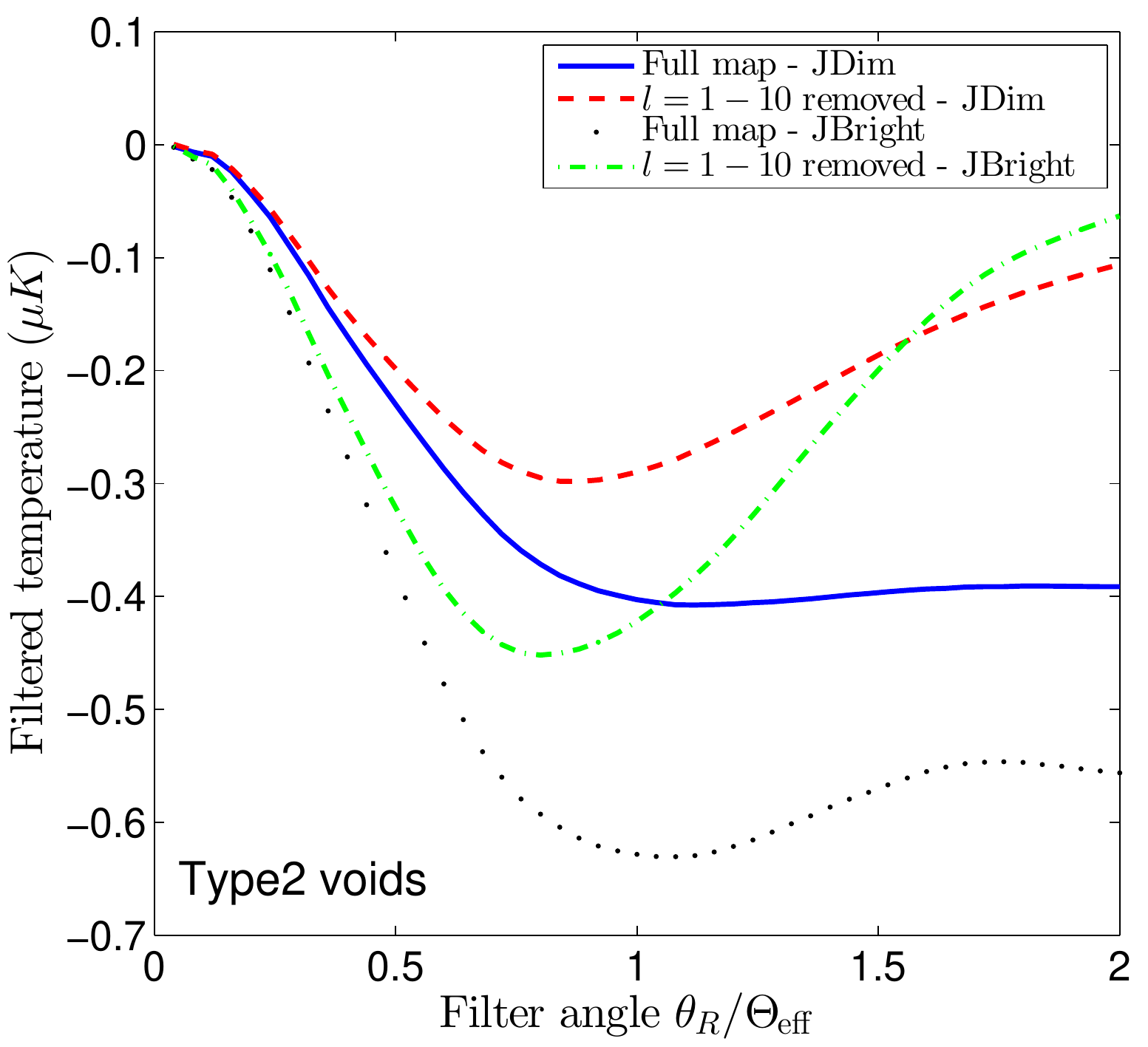}
  \caption{The average filtered ISW temperature anisotropy in Type1 voids (top) and Type2 voids (bottom) as a function of the rescaling weight $\alpha$ in equation \eqref{eq:rescdef}. The filter is defined in equation \eqref{eq:filtdef}. Curves are shown for both JDim and JBright voids found in the Jubilee simulation, with and without large scale modes removed ($C_l=0$ for $l=1-10$).}
  \label{fig:resctype1}
\end{figure}

\begin{figure}
  \centering
 \includegraphics[width=0.45\textwidth]{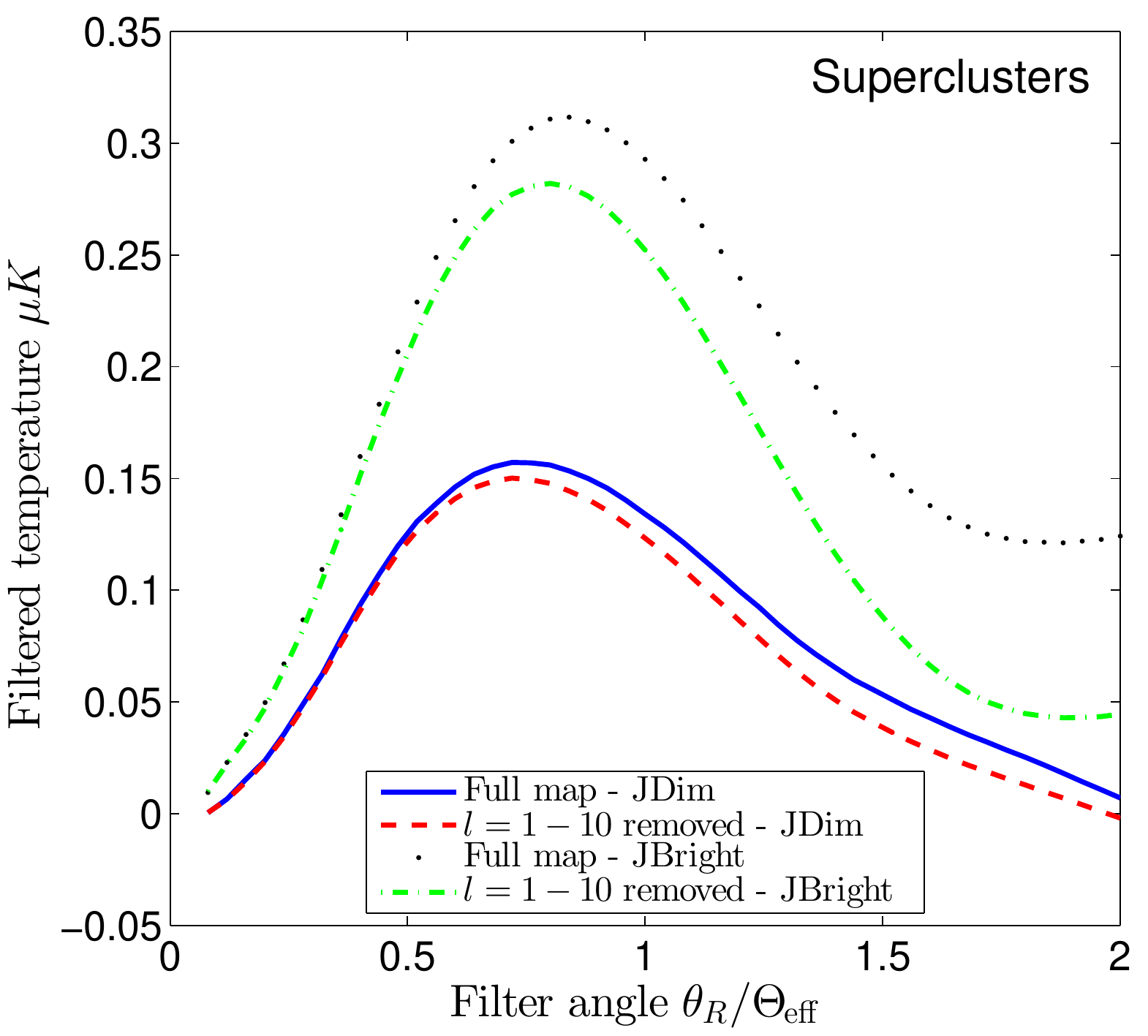} 
  \caption{The average filtered ISW temperature anisotropy in superclusters as a function of the rescaling weight $\alpha$ in equation \eqref{eq:rescdef}. The filter is defined in equation \eqref{eq:filtdef}. Curves are shown for both JDim and JBright superclusters found in the Jubilee simulation, with and without large scale modes removed ($C_l=0$ for $l=1-10$).}
  \label{fig:rescclust}
\end{figure}

Figures \ref{fig:resctype1} and \ref{fig:rescclust} show the average filtered signal as a function of the rescaling weight $\alpha$ for the full stacks of all of each structure type. The optimal rescaling weight shows some dependence on both the structure type definition as well as the properties of the simulation LRG catalogue. The presence of the hot ring seen in Figure~\ref{fig:stackimg} means that Type1 voids show a clear minimum of the filtered temperature at an optimal rescaling weight of $\alpha\simeq0.6$, which is similar to that seen in C13. On the other hand, Type2 voids do not show any clear optimal rescaling weight for the full Jubilee ISW map, with the average filtered temperature instead reaching a plateau at $\alpha \simeq 1$. For superclusters the filtered temperature again shows a pronounced maximum, at $\alpha\sim0.6$ for JDim structures and $\alpha\simeq0.8$ for JBright. In all cases, however, the scale of the maximum average temperature effect is small, and likely to be dwarfed by noise from the primary CMB anisotropies.

It is noticeable that structures from the JBright catalogue consistently have larger average temperature effects than their counterparts from JDim. This is because the sparser distribution of the JBright tracer LRGs means that only larger structures can be detected by {\small ZOBOV}; the mean size of JBright structures is always larger than for corresponding JDim structures and so the photon travel time and ISW temperature shift is also greater. However, JBright catalogues also contain many fewer structures.

Note that it is possible, perhaps even likely, that the optimal rescaling parameter $\alpha$ depends on other properties of the superstructures, e.g. the density ratio $r=\rho_\rmn{link}/\rho_\rmn{min}$ for voids. the detection strategy could therefore be further fine-tuned. However, the \Lam CDM expectation is too small for such fine-tuning to be worthwhile, and given that the signal previously observed by G08---with a constant rescaling---was orders of magnitude larger, it is unlikely that the chance of detecting it, were it to exist in the real data, would be materially affected by the lack of such fine-tuning either.

Figures \ref{fig:resctype1} and \ref{fig:rescclust} also show the effects of removing the largest scale modes ($l\leq10$) from the full ISW map. This was the procedure originally followed by G08 and has been done in all analyses since. The motivation is to reduce `noise' from the largest scales; however it is clear that this subtraction also removes \emph{signal} on these larger scales too. Therefore, whether this is actually optimal strongly depends on the relative size of the removed signal to the removed noise. Note that this conclusion differs from the claim in C13. This is likely to be because the smaller size of the simulation used by C13 meant that these large scale modes are already heavily supressed, or zero.

Given the potentially large effect that can be seen in removing these large scale modes, we choose to include them in our main analysis of the real-world data, and in our further analyses of Jubilee. Clearly, the absolute ISW temperature effect in \Lam CDM is so small that this choice makes no difference. However, in a hypothetical scenario where an ISW-like effect were to show a similar scaling to the \Lam CDM prediction but a much amplified signal strength, the contributions from $l\leq10$ modes would be relevant. In fact, our results taken together with the observation that the original \emph{filtered} G08 signal does not change when these modes are removed \citep{Ilic:2013cn} already constrains such a hypothesis.\footnote{Note that \citet{Ilic:2013cn} do find that the removal of the $l\leq10$ modes changes the absolute temperature anisotropy along the G08 lines of sight, but there is no net contribution to the filtered signal. This is not the case for the Jubilee voids where a net change is observed even in the filtered signal.} However, unless otherwise stated, our main conclusions are unaffected by this choice.

\subsection{Effect of redshift extent of simulation}
\label{sec:redextent}

Although the galaxy catalogues we use to identify superstructures trace the density fluctuations in the Universe, 
the ISW integral is sensitive to fluctuations in the gravitational potential, which extend over much larger scales, as can be seen from the $k^2$ term in Equation~\ref{eq:Poiss}. Therefore it is not clear that the potential fluctuations contributing to the ISW temperature shift of these superstructures are completely contained within the same redshift range as the structures themselves.

Previous studies of the upper bound to the \Lam CDM expectation from simulations \citep[][C13]{Flender:2012wu,HernandezMonteagudo:2012ms} had found that it was much smaller than the G08 observation, but had been open to the criticism that due to their smaller box sizes, the gravitational potential from superstructures could `leak' out of the box, thus accounting for the discrepancy.\footnote{Although note that the theoretical model used by \citet{Nadathur:2011iu} also predicted the upper bound to be similar in size to that of \citet{Flender:2012wu}. There is no ``finiteness of box size'' restriction in a theoretical model.}

\begin{figure}
  \centering
\includegraphics[width=0.45\textwidth]{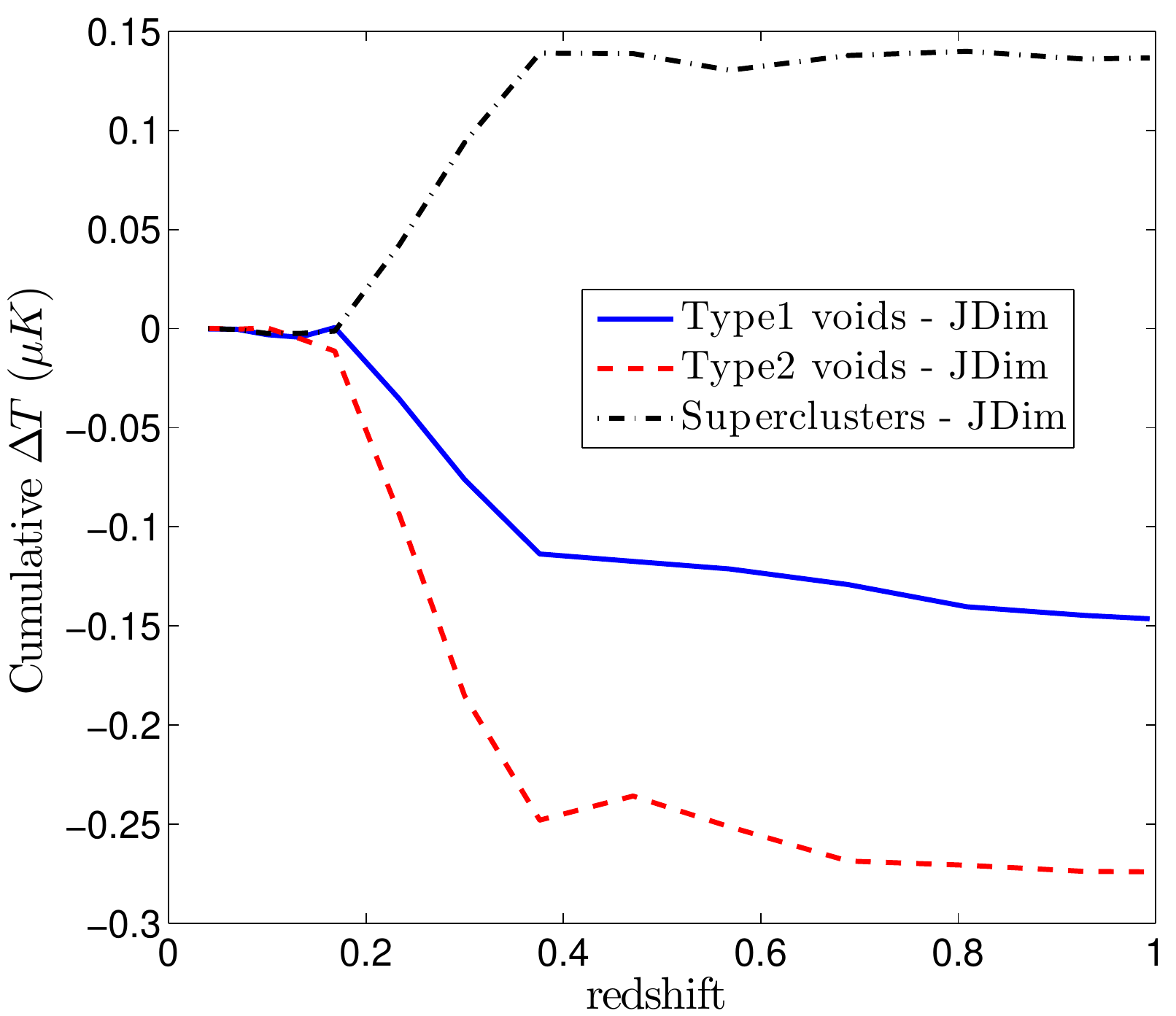} 
  \caption{The cumulative, average, filtered temperature anisotropy from JDim structures found within the Jubilee simulation. The rescaling weight is $\alpha=0.6$ for all curves.}
  \label{fig:redextent}
\end{figure}

The unprecedented size of the Jubilee simulation allows us to settle this question. Taking all the lines of sight to superstructures in the JDim sample (redshift extent $0.16<z<0.36$), we determine the contribution to the total (average, filtered) ISW temperature shift as the photons traverse different redshift ranges. In Figure \ref{fig:redextent} this is shown as the \emph{cumulative} filtered average temperature shift versus the maximum redshift to which the ISW integral is performed. The contribution to the measured $\Delta T$ comes almost entirely from a small redshift range that corresponds closely to the redshift extent of the galaxy distribution, with a plateau at both higher and lower redshifts. This vindicates the assumptions of the previous studies and thus confirms that the G08 measurement is indeed much larger than the \Lam CDM expectation.

\subsection{Parameter dependence of stacked ISW}
\label{sec:selectcut}

The definitions we are using of Type1 and Type2 voids and superclusters were all motivated independently of consideration of their filtered ISW temperature shift. In one sense this is a very good attribute, because it minimises the chance of any accidental \emph{a posteriori} bias caused by making cuts to the catalogue \emph{after} having seen the data. However, we might find that in the real-world some particular selection cut on superstructure properties does strongly improve the observed signal-to-noise ratio. In order to avoid the chances of bias, we need to perform a blind analysis of the effects of possible cuts on these properties in the simulation before turning to actual observation. The parameters we examine for such cuts are the minimum density $\rho_\rmn{min}$ of voids (maximum $\rho_\rmn{max}$ for superclusters), the effective radius $R_\rmn{eff}$ and the density ratio $r$.

To do this we split the superstructure samples into bins according to the value of each parameter, and measure the mean filtered temperature shift for each bin, and estimate the standard error in this mean from the standard deviation.\footnote{Note however that for small numbers of structures in the bin, even if the population distribution is Gaussian, the errors will follow Student's t-distribution, which has wider tails.} The binning is performed such that the bins have equal numbers of structures, to avoid the problem of empty or nearly empty bins. The only bin with unequal numbers is the final bin, which always has additional structures. Parameter values for $\rho_\rmn{min/max}$, $R_\rmn{eff}$ and $r$ for any bin are always taken to be the central point of the bin range. The rescaling weight is kept fixed throughout at $\alpha=0.6$. The results for different parameter dependencies are shown in Figures~\ref{fig:mindens},~\ref{fig:effrad} and \ref{fig:densrat}.

\subsubsection{Density Extremum}
\label{sec:mindens}
\begin{figure}
  \centering
  \includegraphics[width=0.45\textwidth]{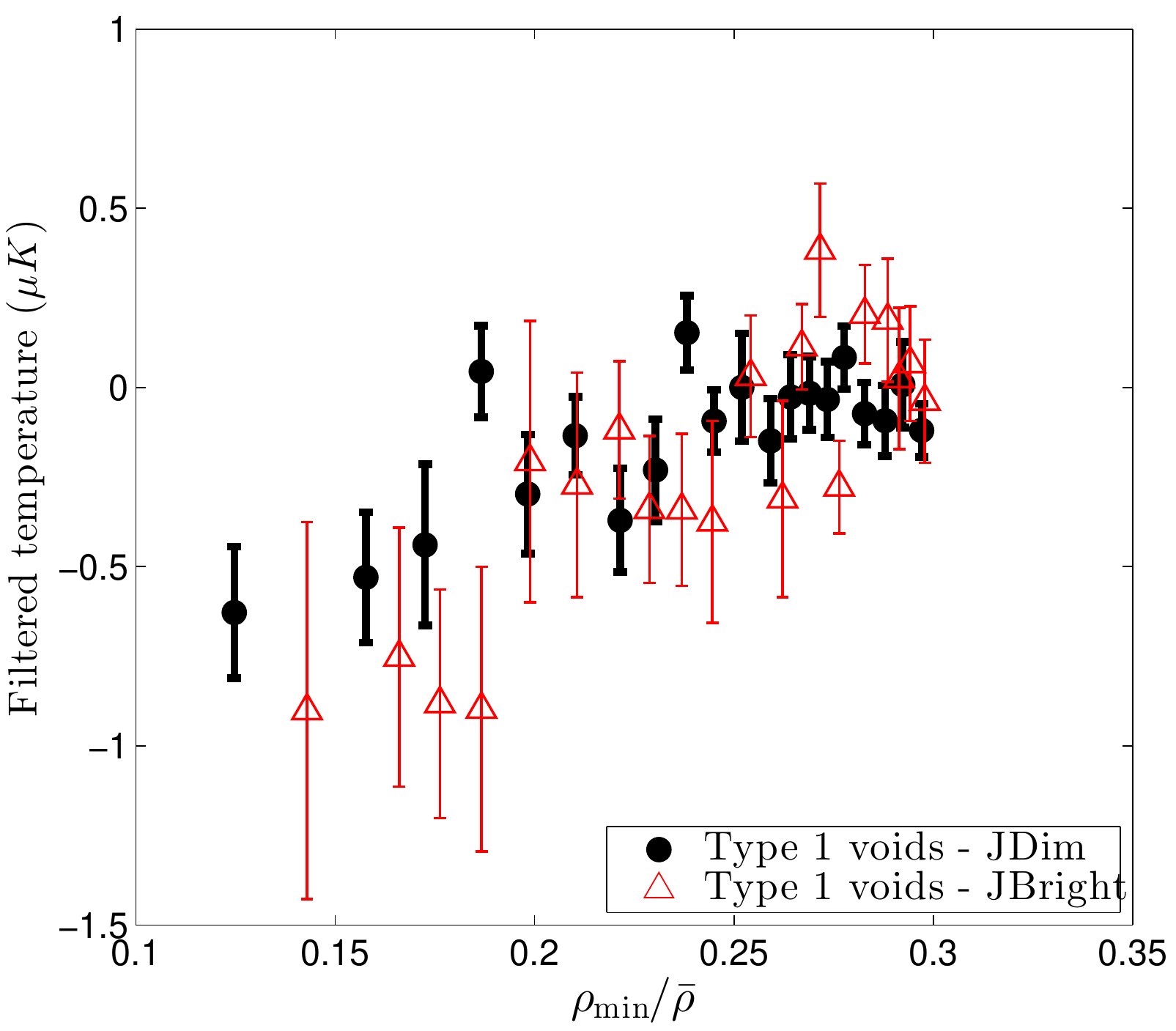}
  \includegraphics[width=0.45\textwidth]{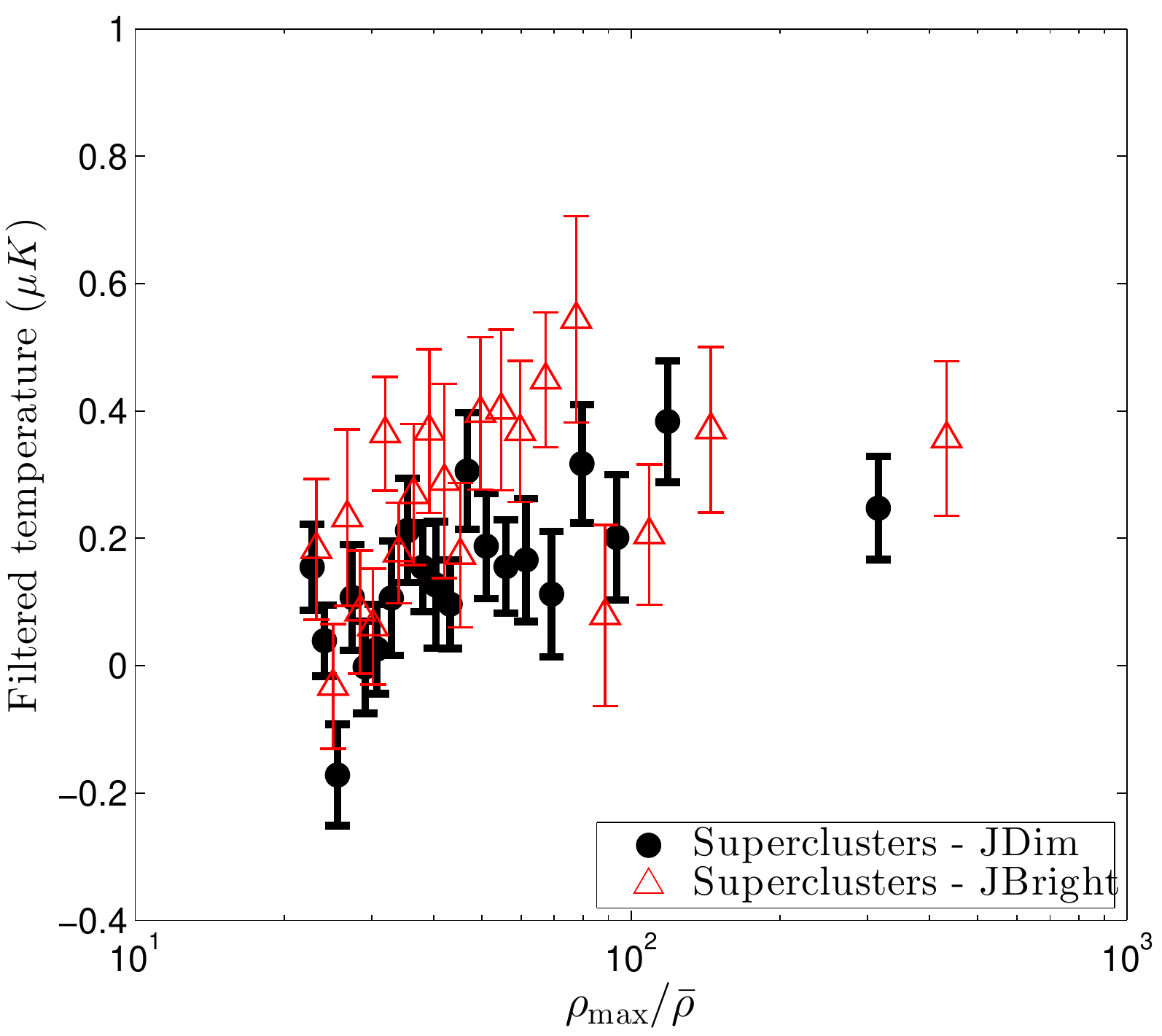}
  \caption{The average, filtered, ISW temperature anisotropy caused by Type1 voids (top) and superclusters (bottom) found in the Jubilee simuation as a function of minimum density and maximum density respectively.}
  \label{fig:mindens}
\end{figure}
Figure \ref{fig:mindens} shows the dependence of $\Delta T$ on $\rho_{\rm min}$ ($\rho_\rmn{max}$) for Type1 voids (superclusters). There is a clear trend towards larger temperature shifts from voids with more extreme minimum densities. This is entirely expected: larger density fluctuations should correspond to larger potential fluctuations (Equation~\ref{eq:Poiss}) and therefore larger temperature shifts. There is also a trend for superclusters, though its precise shape is less clear (note the logarithmic axis).

Note that trend lines through the JDim and JBright Type1 void data points would reach $\Delta T=0$ very close to the density cutoff $\rho_\rmn{min}/\overline{\rho}=0.3$ applied in the superstructure selection (Section~\ref{sec:Jubstruc}) to distinguish genuine structures from Poisson noise. At least for the voids, this suggests that our \emph{a priori} choice of selection criteria was well motivated. We note that our void definition differs in this respect from that used by C13, who required only that $\rho_\rmn{min}/\overline{\rho}\leq1$ but subsequently found that further cuts on $R_\rmn{eff}$ were required to ensure that their sample of voids corresponded to a negative average temperature shift. This suggests that the `void-in-cloud' problem that they refer to is simply avoided by requiring a minimum underdensity in the definition of a void.

Note also that this has important implications for the stacking studies of \citet{Ilic:2013cn,Planck:ISW}, who used a sample of voids from \citet{Sutter:2012wh} for many of which the minimum density was in fact $\rho_\rmn{min}>\overline{\rho}$ \citep[NH14;][]{NH:2013b}, i.e. they were not voids at all and would not be expected to cause a negative ISW temperature shift.

\subsubsection{Effective radius}
\label{sec:effrad}

\begin{figure}
  \centering
   \includegraphics[width=0.45\textwidth]{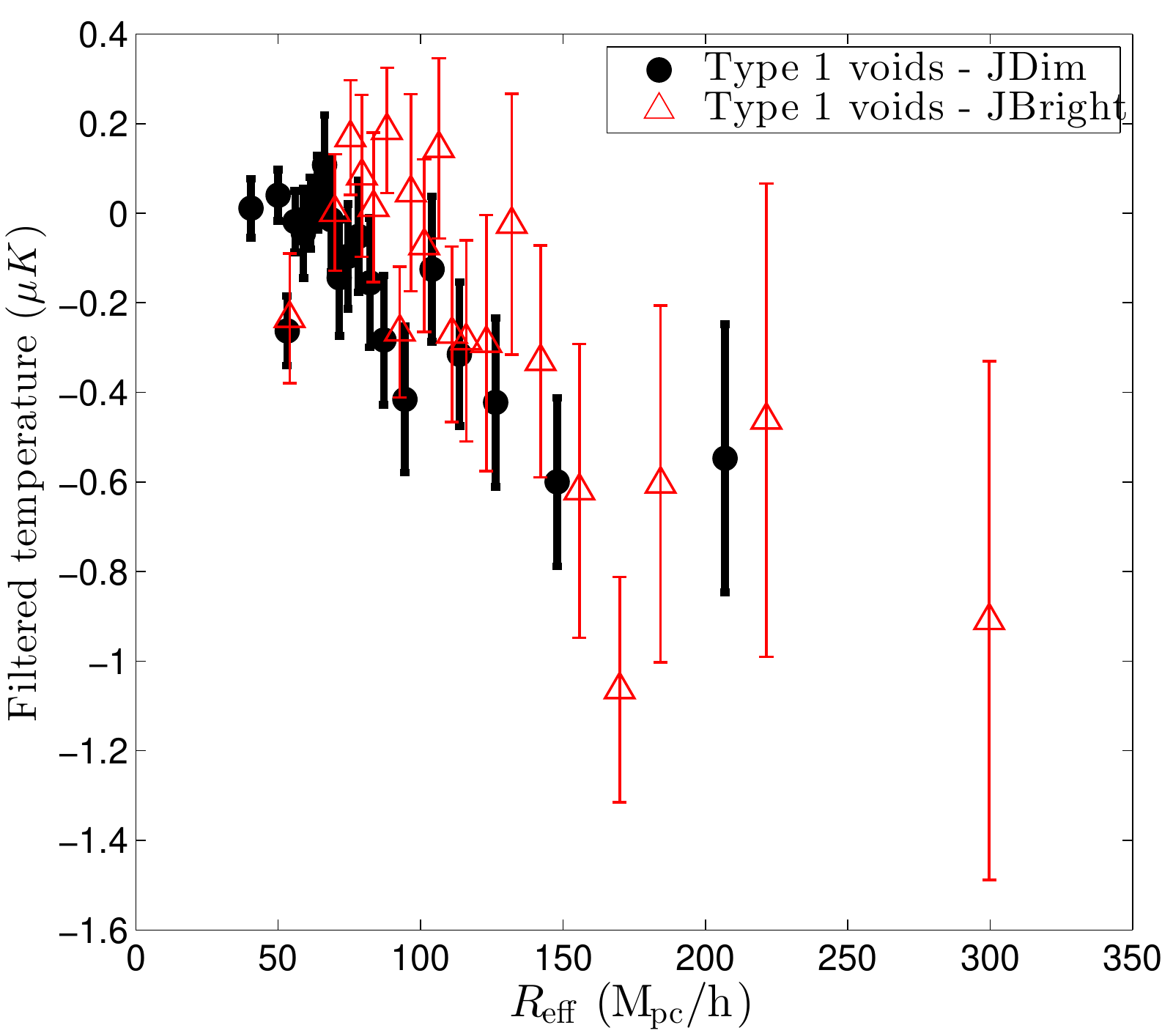}
   \includegraphics[width=0.45\textwidth]{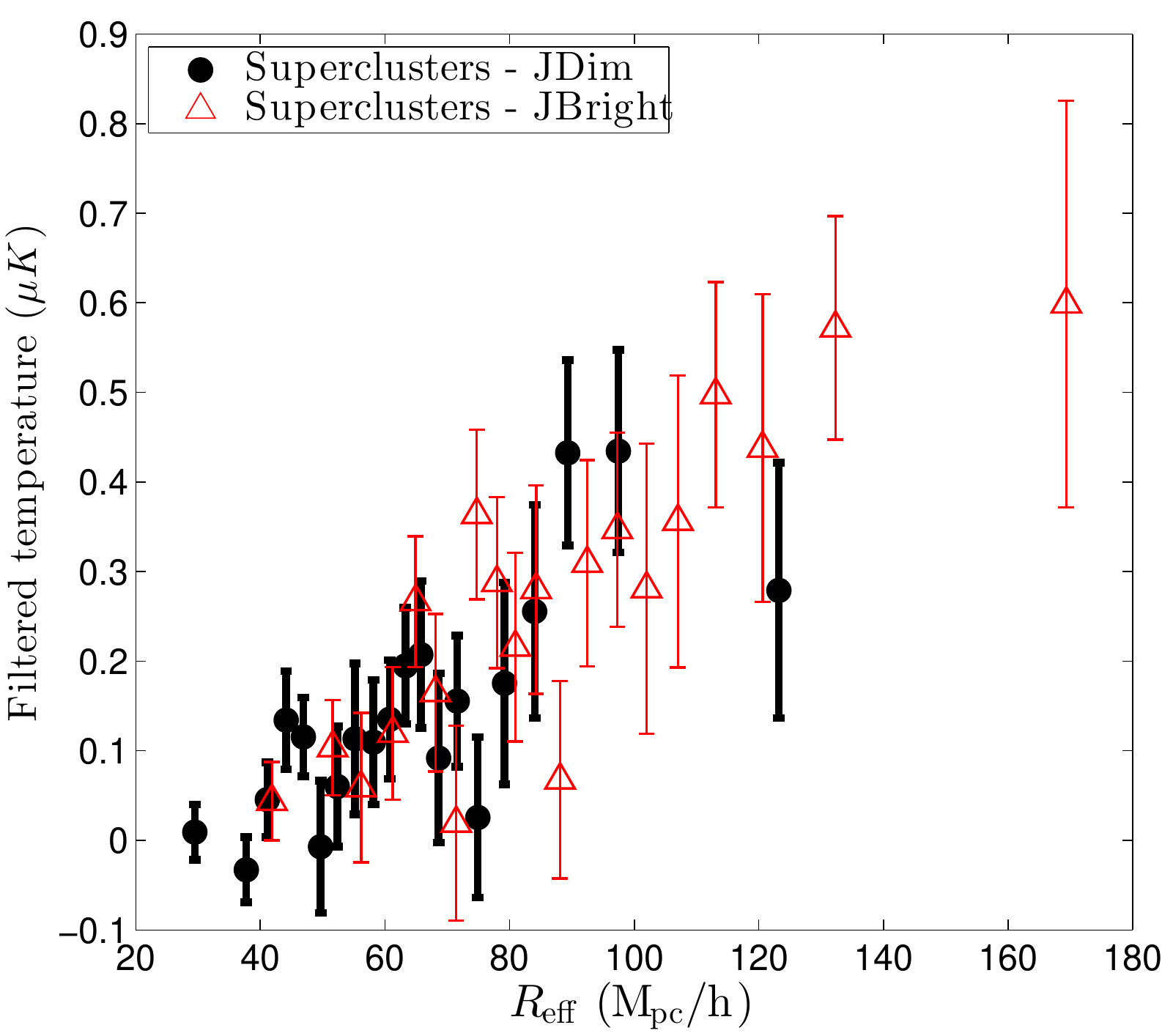}
  \caption{The average, filtered, ISW temperature anisotropy caused by Type1 voids (top) and superclusters (bottom) found in the Jubilee simuation as a function of the effective radius of the superstructure.}
  \label{fig:effrad}
\end{figure}

Figure \ref{fig:effrad} shows the dependence of $\Delta T$ on $R_{\rm eff}$. Unsurprisingly, both voids and superclusters show a clear trend towards larger temperature shifts signal for more extreme structures (here meaning larger radius). This is because the photon travel time through larger structures is longer, thus increasing the contribution to the ISW integral of Equation~\ref{eq:ISW}.

Note that for the same value of $R_{\rm eff}$ structures from JBright do not give systematically larger signals than ones from JDim (indeed if anything they give smaller signals). However JBright structures are on average clearly larger. This confirms that the trends seen in Figures~\ref{fig:resctype1},~\ref{fig:rescclust} and \ref{fig:mindens} towards larger temperature shifts from JBright structures are primarily due to their larger sizes.

\subsubsection{Density ratio}
\label{sec:densrat}

\begin{figure}
  \centering
   \includegraphics[width=0.45\textwidth]{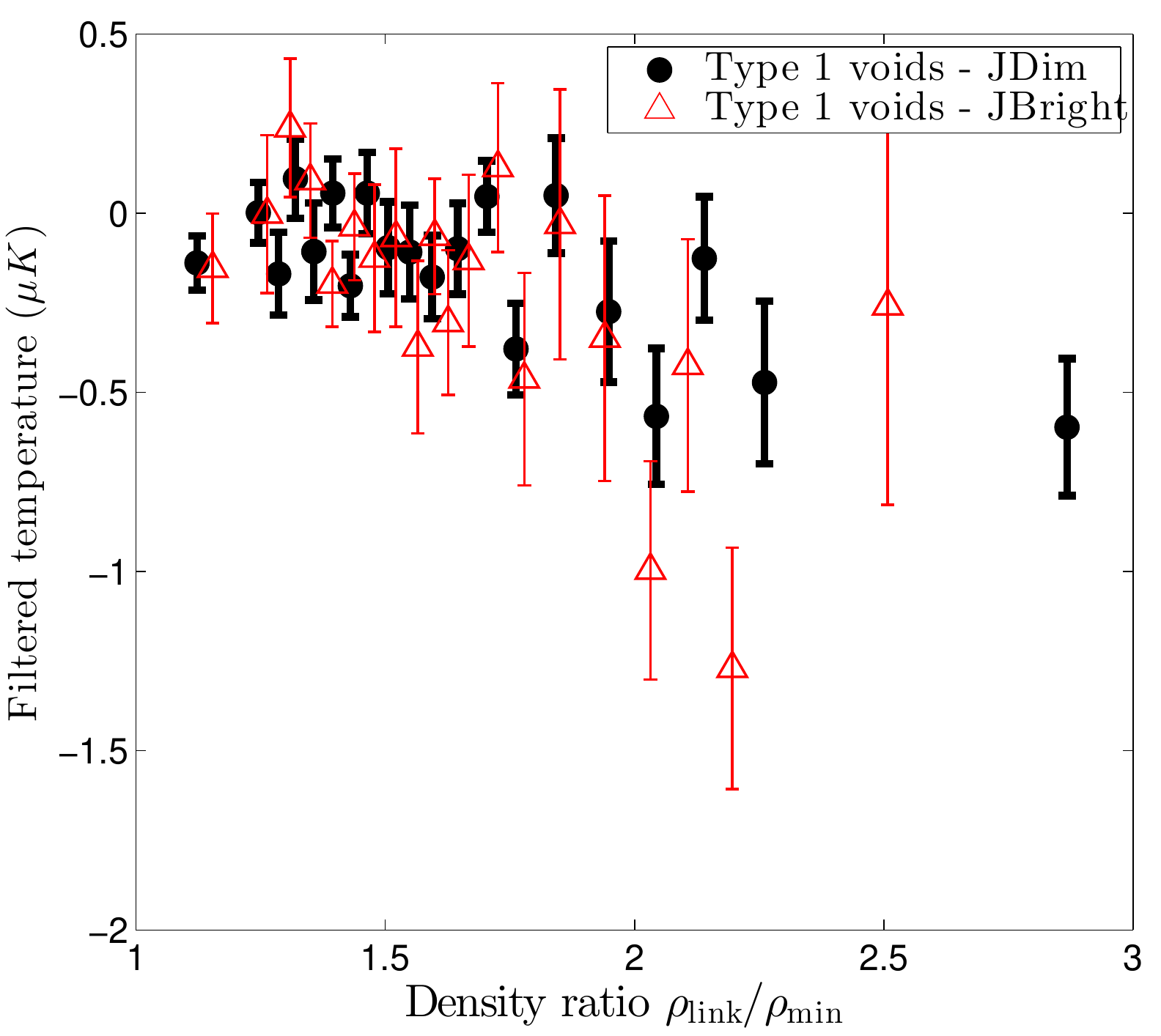} 
   \includegraphics[width=0.45\textwidth]{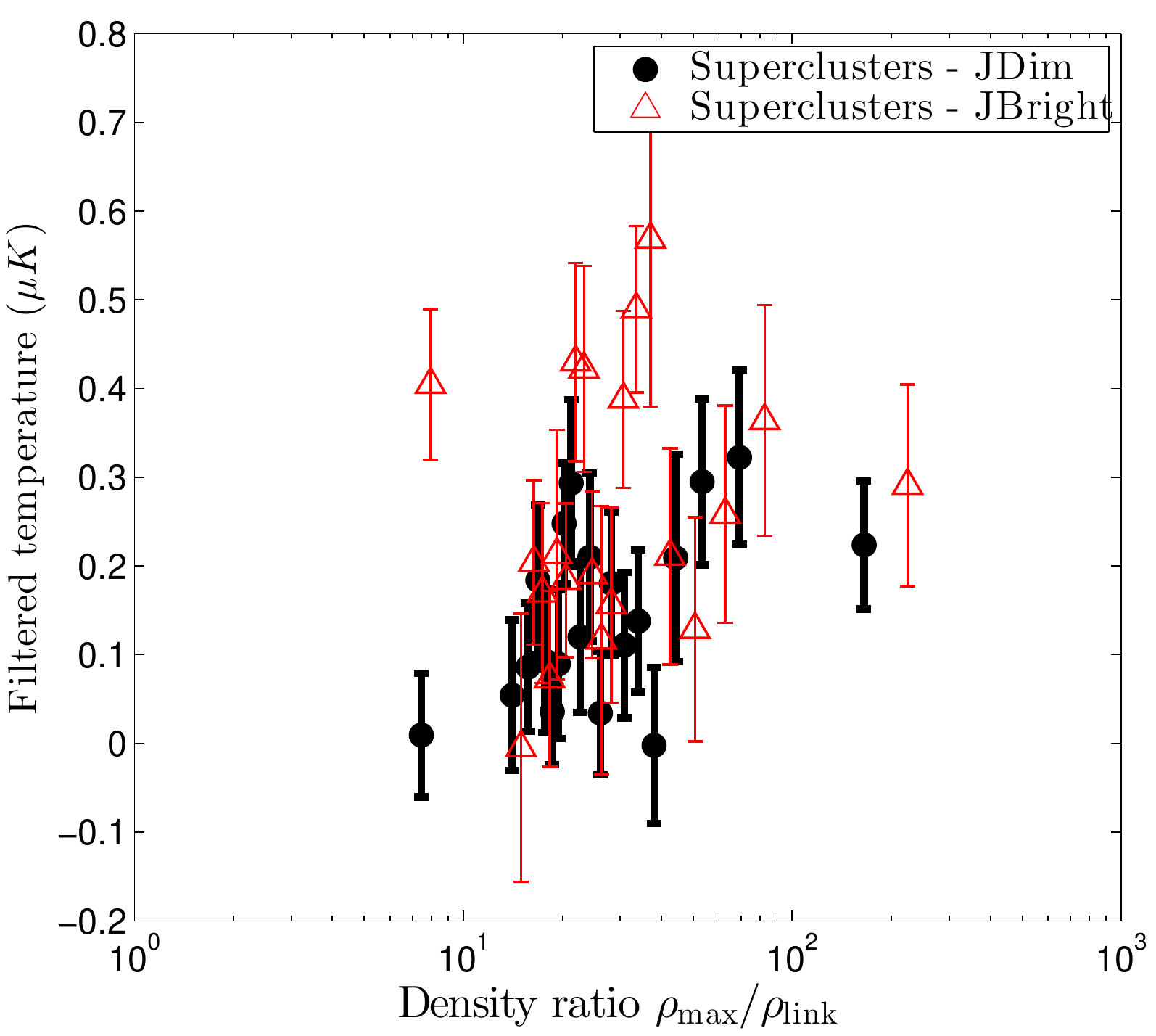} 
  \caption{The average, filtered, ISW temperature anisotropy caused by Type1 voids (top) and superclusters (bottom) found in the Jubilee simuation as a function of the density ratio parameter, $r$.}
  \label{fig:densrat}
\end{figure}

Finally, Figure \ref{fig:densrat} shows the dependence of $\Delta T$ on the density ratio $r$ (defined for voids as $\rho_\rmn{link}/\rho_\rmn{min}$ and for superclusters as $\rho_\rmn{max}/\rho_\rmn{link}$). There is a trend towards increasing signal with larger $r$, as would be expected given the trends with $\rho_\rmn{min/max}$ though the dependence on $r$ is relatively weak, particularly for voids with $r\lesssim2$. However, the density ratio is less discriminating than the structure radius $R_\rmn{eff}$.  

\subsection{Summary of \Lam CDM expectation}
\label{sec:ISWsimsum}

The results of this section vindicate earlier conservative estimates of the maximum possible stacked ISW signal in \Lam CDM \citep{Nadathur:2011iu, Flender:2012wu,HernandezMonteagudo:2012ms} which claimed a discrepancy between this expectation and the G08 observation. We do not find any subset of superstructures in Jubilee, based on any choice of density, size or density ratio, that can explain a stacked signal amplitude of $\gtrsim1\;\umu $K. This is an order of magnitude smaller than the G08 measurement, so even accounting for the slightly different redshift range of the G08 LRG sample, it is clear that the observed signal cannot be due to the ISW effect in a \Lam CDM universe. Finally, reproducing figures \ref{fig:resctype1} -- \ref{fig:densrat} with a fixed filter radius produces an even smaller signal than the ones presented for a rescaled filter radius.

Our simulation results also show noticeable trends between the $\rho_\rmn{min/max}$, $R_\rmn{eff}$ and $r$ values for superstructures and their stacked ISW temperatures. Any signal found in the real data would be expected to follow these trends if it were due to the late-time decay of potentials (even in a model other than \Lam CDM). Conversely, if a signal seen in the real world does not show such trends, this would be strongly suggestive of a systematic error or random fluctuation.

%----------Next section------------%

\section{The stacked ISW signal in SDSS data}
\label{sec:ISWdata}

We now turn to the catalogues of superstructures from SDSS data, and look for evidence of their effects on the Planck CMB maps. It should be kept in mind that, if \Lam CDM is correct, we would not expect to see anything statistically significant. If we examine the data in a sufficient number of different ways \emph{a posteriori}, a $2-3\sigma$ significant signal may easily be seen in one of them. But unless this signal follows the trends with parameters described above, or appears when we use precisely the same methodology as was used in G08, this would not constitute strong evidence of any new physics.

\subsection{Procedure}
\label{sec:proced}

We use the Planck SMICA CMB map \citep{Planck:Overview} and the conservative Union (U74) mask to remove contamination from the galaxy and point sources. From this map we extract patches around the lines of sight of superstructures in the SDSS catalogues and apply the compensated top-hat filter as described in Equation \eqref{eq:filtdef}. Unless explicitly stated, we do not remove modes $l\leq10$, but this makes no difference to our conclusions. in performing the filtering, we treat masked pixels as pixels with zero temperature and take the average over the full discs.

To estimate the noise in our measurements we keep the same distribution of scaled filter radii for superstructures, but randomise the lines of sight within the SDSS window before calculating the average temperature. We repeat this process 1000 times and take the standard deviation of these 1000 samples to be the uncertainty in the measurement.

This method is not entirely satisfactory because it is unable to capture any effects that might arise due to the correlation of the lines of sight. This is potentially important because, in a single large under-dense (overdense) region of a survey, {\small ZOBOV} may report multiple voids (superclusters). A completely random choice of directions may therefore show less clustering than the specific lines of sight chosen by {\small ZOBOV}. The variance of the less clustered set of directions will be smaller, so this method overestimates the true signal-to-noise (S/N). However, as we claim only a null detection, this is unimportant.

An alternative method to estimate the uncertainty is to generate a large number of sample maps and to examine the stacked signal using the same directions and filter sizes on each of these sample maps. This accounts for the correlation between lines of sight; however, in the event of a statistically significant S/N it is unable to distinguish between the possibilities that the lines of sight to superstructures within the SDSS window are special, or that the SDSS window itself is special. 

The estimates of the variance about the mean obtained from these two different methods are in fact very similar \citep[G08;][C13]{Ilic:2013cn}. However, by restricting our randomised lines of sight to be within the SDSS window we find that the \emph{mean} within this window is itself generally significantly \emph{non}-zero. For example, for the set of structure sizes associated with superclusters from \emph{lrgdim}, and a rescaling weight of $\alpha=0.6$, we find the mean of the 1000 realizations of the $\overline{\Delta T}$ measurement for constrained random directions is $0.57\;\umu$K, where the expected error in the mean is only $\sim10^{-3}\;\umu$K. This mean offset is dependent on both the number of structures in a given subset and the rescaling weight $\alpha$. It is a characteristic of the CMB in the region of the SDSS window; since our superstructure directions must necessarily lie within this window, the non-zero mean should be taken into account.

\subsection{Results}
\label{sec:results}

We start by examining the full Type1 and Type2 void catalogues and  supercluster catalogues for the \emph{lrgdim} and \emph{lrgbright} galaxy samples from NH14. These are two galaxy samples that our mock LRG catalogues in Jubilee were designed to replicate. 

For our first pass, we set a fixed rescaling weight $\alpha=0.6$ for all voids and superclusters. This is the optimal weight found for Type1 voids and superclusters in Jubilee, and also matches the optimal weighting found by C13. With this rescaling weight, for superstructures in the \emph{lrgdim} sample we find $\overline{\Delta T} = 0.14 \pm 2.8\; \umu$K for the 70 Type1 voids, $\overline{\Delta T} = 2.6 \pm 5.2\; \umu$K for the 19 Type2 voids and $\overline{\Delta T} = 2.05 \pm 1.9\; \umu$K for the 196 superclusters. In \emph{lrgbright} we find $\overline{\Delta T} = -2.6 \pm 5.4\; \umu$K for the 13 Type1 voids, $\overline{\Delta T} = -50 \pm 23\;\umu$K for the only Type2 void and $\overline{\Delta T} = -4.0 \pm 3.8\;\umu$K for the 39 superclusters.

Clearly, none of these results have any statistical significance, and this fact is unchanged when we account for the small systematic shifts due to the non-zero mean expectations within the SDSS window. However, we stress that such a result is \emph{precisely} what would be expected in \Lam CDM, since the expected signal is so small that it is dominated by the noise from the primary anisotropies. Extending the superstructure catalogues to include those found in the main galaxy samples \emph{dim1} to \emph{bright2} also did not produce any significant S/N detection. There are many fewer structures found in SDSS than in Jubilee, which is an obvious consequence of the smaller survey window and the effect of the mask. This leads to a relative lack of statistical power in the SDSS samples. Nevertheless our Type1 lrgdim samples still contain more structures than were used in the original anomalous G08 measurement, so if that effect we real we should be able to see it. It is also the case that, in a $\Lambda$CDM cosmology, even the statistical power available in Jubilee is not enough to overcome noise from primary CMB anisotropies -- the signal really should be unobservable
\begin{figure}
  \centering
     \includegraphics[width=0.44\textwidth]{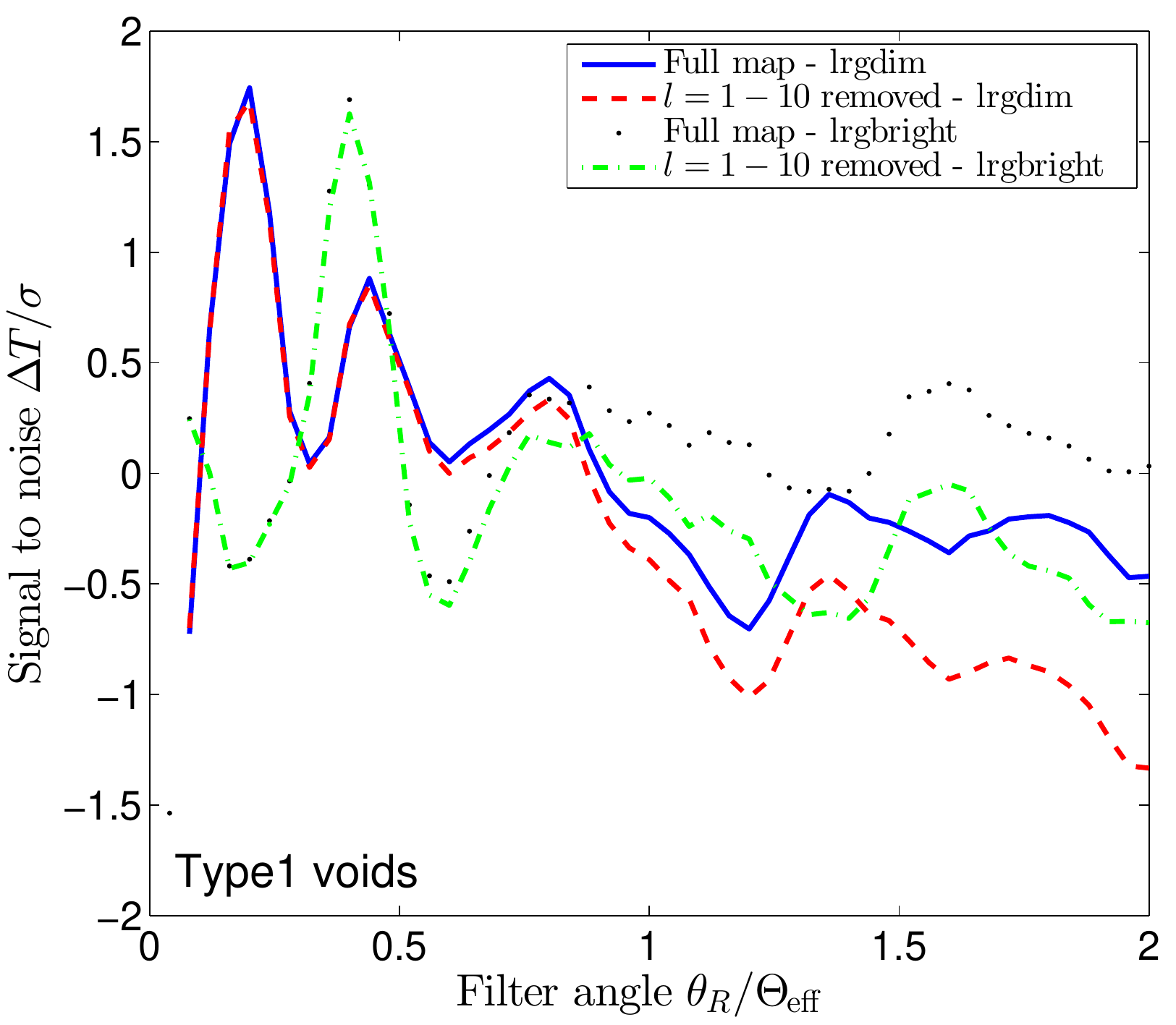} 
     \includegraphics[width=0.44\textwidth]{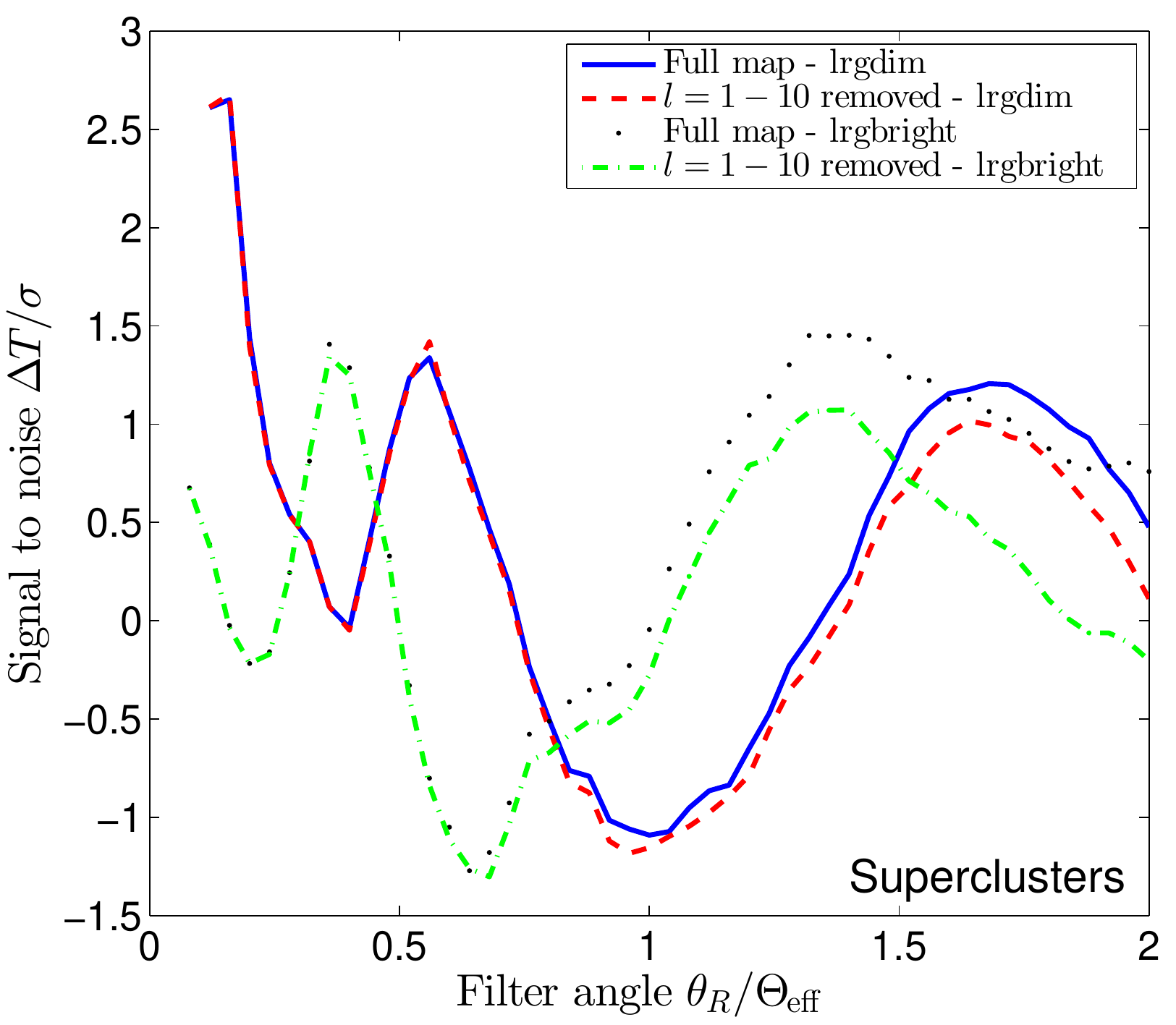} 
  \caption{The average filtered ISW temperature anisotropy in Type1 voids (top) and superclusters (bottom) as a function of the rescaling weight $\alpha$ in equation \eqref{eq:rescdef}. The filter is defined in equation \eqref{eq:filtdef}. Curves are shown for both \emph{lrgdim} and \emph{lrgbright} structures found in \citet{Nadathur:2014a} (NH14), with and without large scale modes removed ($C_l=0$ for $l\leq 10$).}
  \label{fig:diffresc}
\end{figure}

Figure \ref{fig:diffresc} shows the S/N behaviour for Type1 voids and superclusters as the rescaling weight is changed around the optimal value determined from simulation. There is clearly no coherent signal around the expected optimal rescaling weights, and the S/N does not exceed $>2.5\sigma$ at \emph{any} rescaling weight. Repeating the parameter-based searches shown in Figures~\ref{fig:mindens} - \ref{fig:densrat} for this data also fails to reveal any significant S/N. Again, however, this exactly matches the \Lam CDM expectation of a null detection.

\begin{figure}
  \centering
       \includegraphics[width=0.45\textwidth]{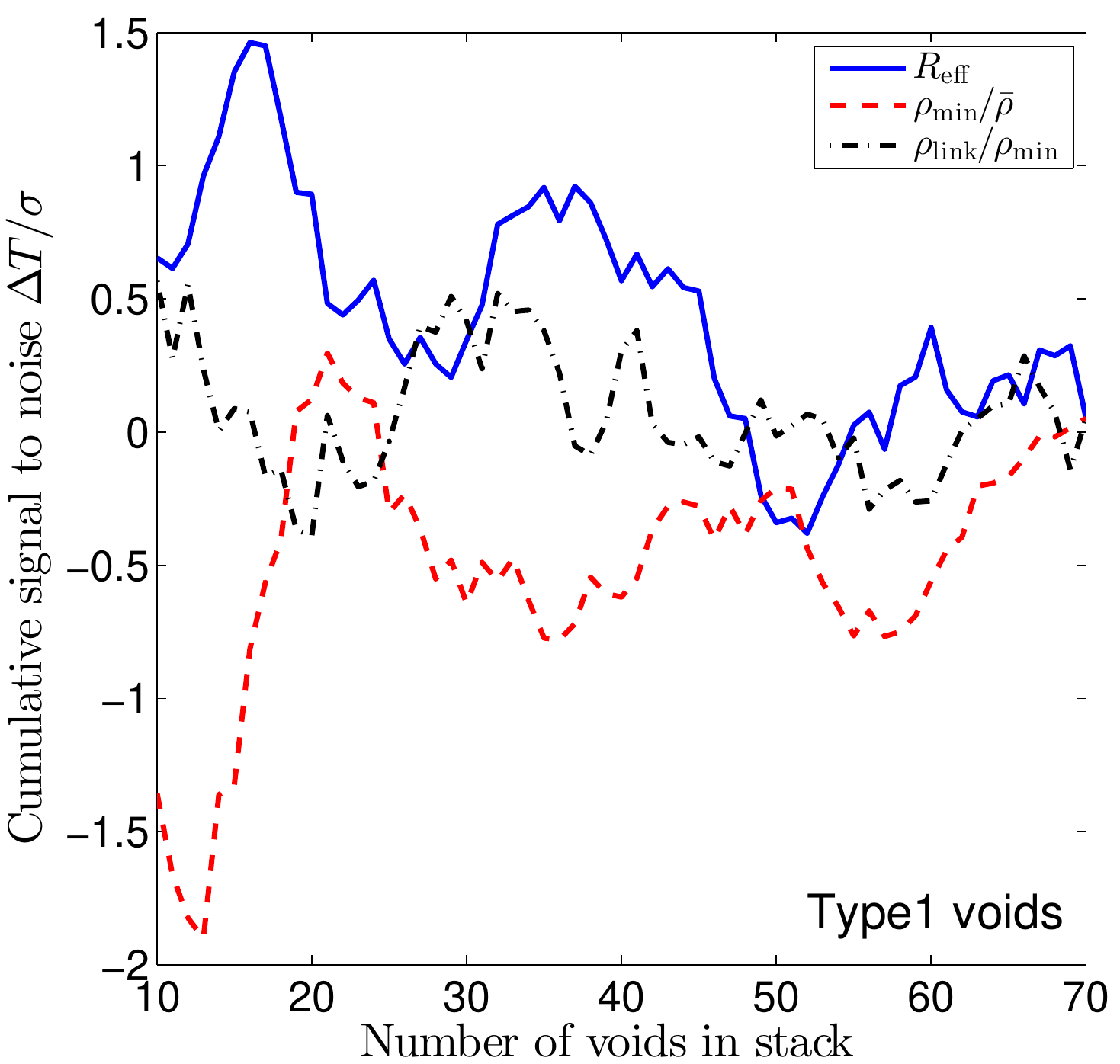} 
       \includegraphics[width=0.45\textwidth]{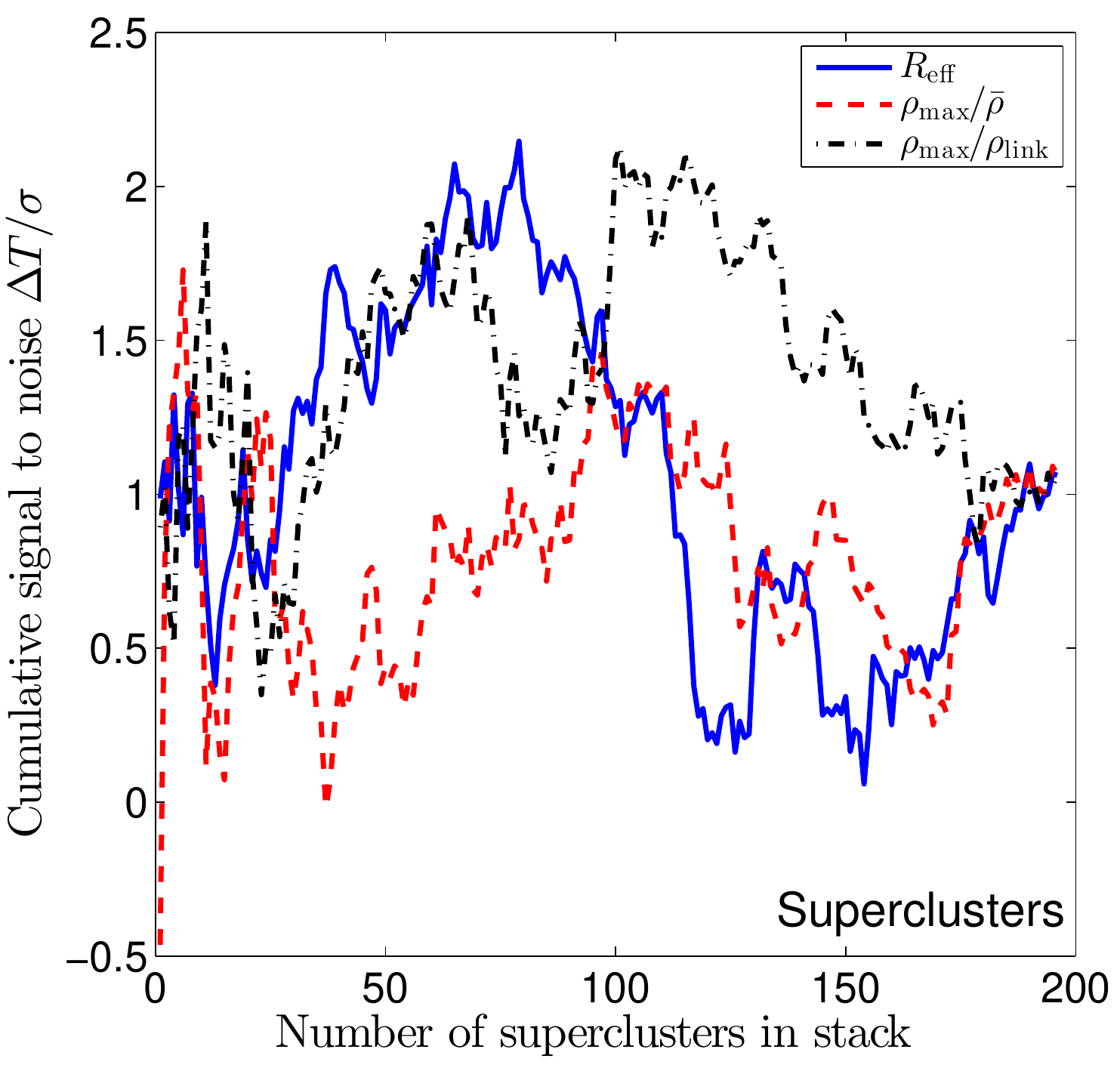} 
  \caption{The cumulative signal-to-noise ratio in Type1 voids (top) and superclusters (bottom) when only the most extreme $N$ structures from \citet{Nadathur:2014a} (NH14) are included in the stack. `Extremeness' is defined here by largest effective radius, largest density ratio parameter or smallest (largest) minimum (maximum) density for voids (superclusters) for the three different curves.}
  \label{fig:LEE}
\end{figure}

However, the G08 observation was based on selecting the 50 voids and 50 superclusters with the largest values of the density ratio parameter $r$, and it is possible that such a cut would give a larger signal. To attempt to mimic this selection criterion as closely as possible, we sort all superstructures in order of their $r$ values, and study the behaviour of the cumulative S/N (with $\alpha=0.6$) as successively larger numbers of structures are added to the stack in this order. For good measure, we also perform the same test after ranking the superstructures by their $\rho_\rmn{min/max}$ and $R_\rmn{eff}$ values. The results  are shown in Figure \ref{fig:LEE} for Type1 voids and superclusters. There is no particularly significant result, for any subset of superstructures, in either case. 

The lack of a significant signal in this sample, particularly when ranking by $r$, leads us to conclude that the signal seen in G08 does not exist in these independent galaxy samples. Although we have shown results using a rescaled filter radius, we arrive at the same conclusion if we instead use a fixed filter radius. There are still some ways in which way have not exactly reproduced the methodology of G08---e.g., the MegaZ LRG sample used for that analysis was photometrically selected and had a higher mean redshift than our spectroscopic samples---which could perhaps be argued to reconcile our null detection. For instance, the hypothesised non-standard physics which explains the amplitude of G08 result might be strongly redshift-dependent. To conclusively exclude even this possibility would require data from another galaxy survey covering the same redshift interval but a different portion of the sky, such as the Dark Energy Survey \citep{DES} may provide in a few years' time.

In our opinion, however, such explanations are somewhat far-fetched. Instead it is more likely that our null result calls into question the physical significance of the G08 detection. We note again in this context that previous analyses have found that when the G08 catalogue is re-analysed using rescaled filter radii based on the size of each void, rather than a fixed size filter for all voids, the observed signal \emph{decreases} in magnitude \citep{Ilic:2013cn}. 

\section{Comparison with previous work}
\label{sec:prevwork}

Some previous studies \citep[][C13]{Ilic:2013cn,Planck:ISW} have looked for the stacked ISW signal for \emph{voids alone} using the same galaxy samples as we have used in this paper, with both WMAP 9-year \citep{Hinshaw:2012aka} and Planck CMB data.

In \citet{Ilic:2013cn,Planck:ISW} no significant detection was claimed. Unfortunately, although the analysis in these works is correct in itself, both these studies made use of a flawed catalogue of voids, as explained in detail elsewhere \citep[][NH14]{NH:2013b}. This makes it difficult to draw any conclusions from their results without reproducing their analysis using a robust catalogue.

On the other hand, the authors of C13 use their own, more robust, void catalogue and $N$-body ISW simulations. These simulations had a smaller box size and used the halo distribution directly rather than simulating a tracer galaxy population. They also used a different selection criterion for voids than we do, preferring to apply a very loose cut on $\rho_\rmn{min}$ in the first instance, but subsequently applying a much stricter cut on the void size $R_\rmn{eff}$ after calibrating their expectations for the $\overline{\Delta T}$ signal from simulations. Although $\rho_\rmn{min}$ and $R_\rmn{eff}$ are quite strongly correlated, this correlation is not perfect, so their final sample of voids differs somewhat from ours. Nevertheless, they found a maximum of the simulated temperature signal at $\alpha=0.6$, similar to our result.

Using their alternative void catalogue from SDSS data and the Planck SMICA CMB map with this value of $\alpha$, they find a filtered temperature of $\overline{\Delta T}\simeq -2.9 \umu$K. This has the correct sign for an ISW-like effect, and a statistical significance of $\sim 2.3\sigma$ (by coincidence, in this case the effect of the non-zero mean in the SDSS window discussed in section \ref{sec:proced} is small and does not affect this result).
\begin{figure}
  \centering
       \includegraphics[width=0.45\textwidth]{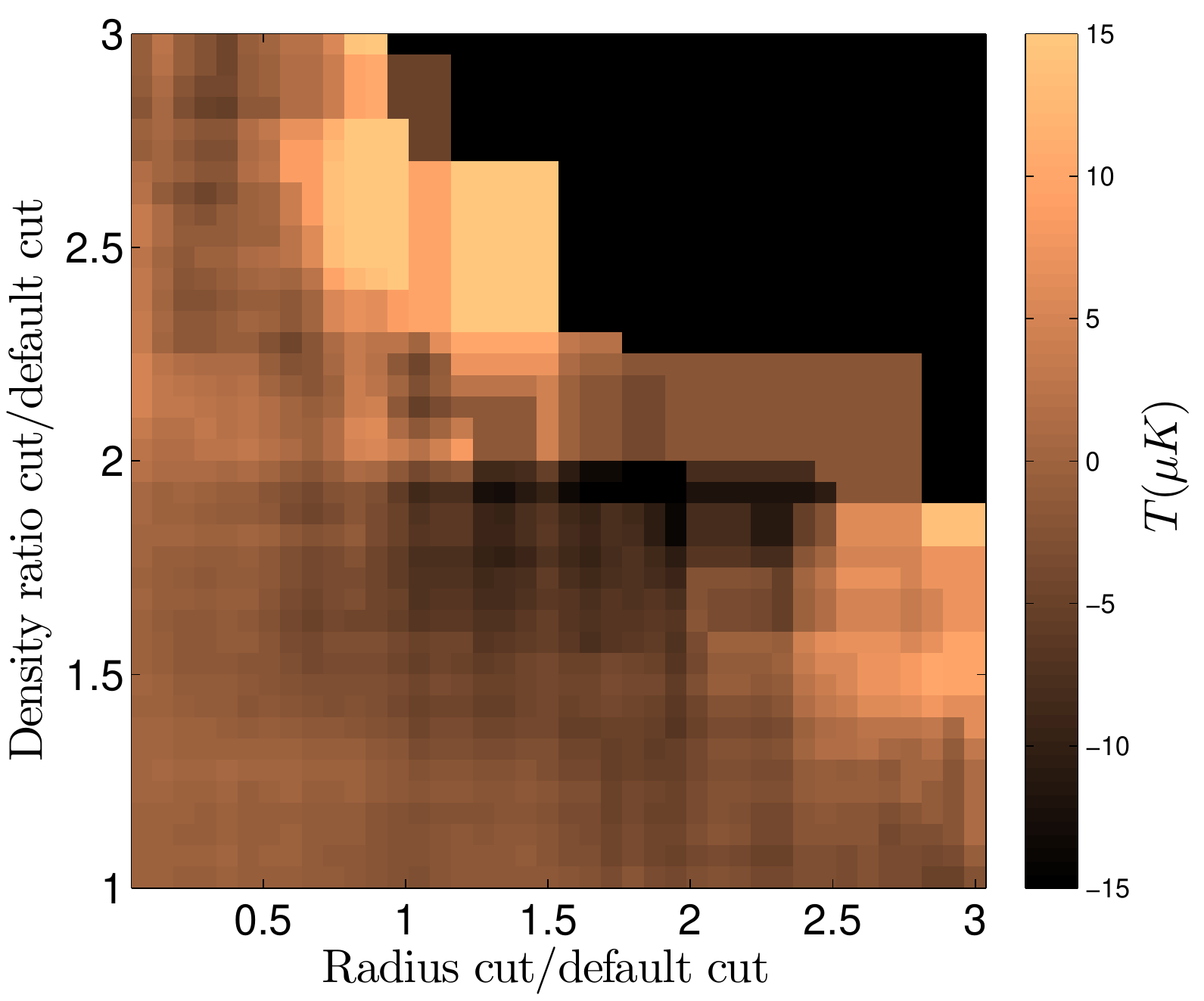} 
  \caption{The average, filtered temperature anisotropy in (real-world) voids from \citet{Cai:2013ik} (C13) as a function of cuts on radius and density ratio parameter. Neighbouring data points are not independent and contain mostly the same voids.}
  \label{fig:Caiheat}
\end{figure}

Firstly, we observe that the C13 voids are detected using the so-called `redshift space' coordinate system, in which the radial coordinate of a galaxy is not its comoving distance but is linearly proportional to its redshift.\footnote{Note that this is not the same `redshift space' as the one referred to in discussion of physical effects such as redshift space distortions.} This is not the same comoving coordinate system as we have used, nor the same as C13 themselves apply in the simulations they use for calibration. The primary effect of this coordinate system is to increase measured void volumes---and therefore $R_\rmn{eff}$ values---in a redshift-dependent manner, though it also distorts their shapes by stretching along the line of sight and may affect the detection of voids itself (NH14). As the ISW effect is sensitive to the gravitational potential and thus to density fluctuations in physical (comoving) coordinates, one would ordinarily expect the stacked signal to be larger for a catalogue found using comoving coordinates.
\begin{figure}
  \centering
       \includegraphics[width=0.45\textwidth]{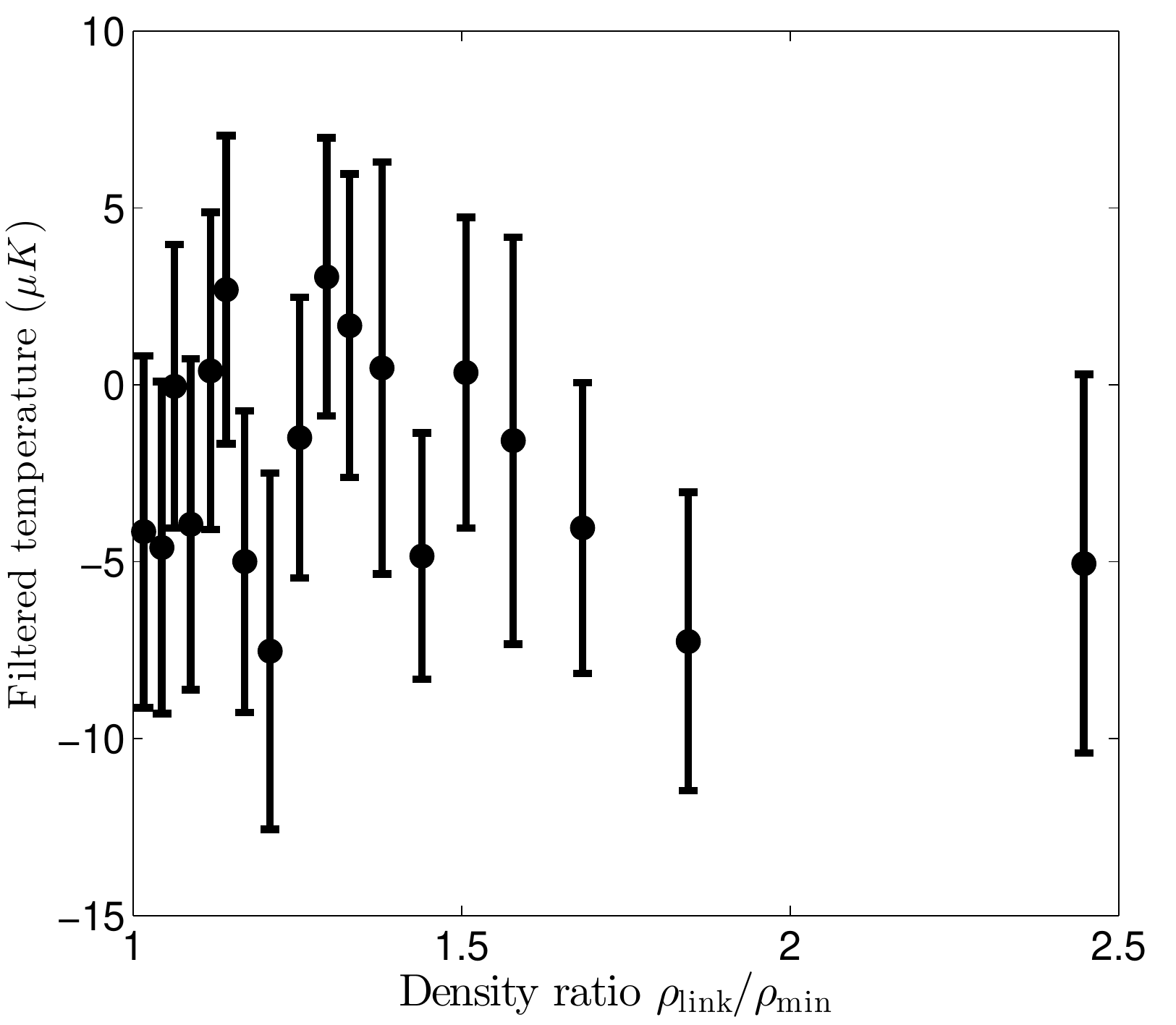} 
       \includegraphics[width=0.45\textwidth]{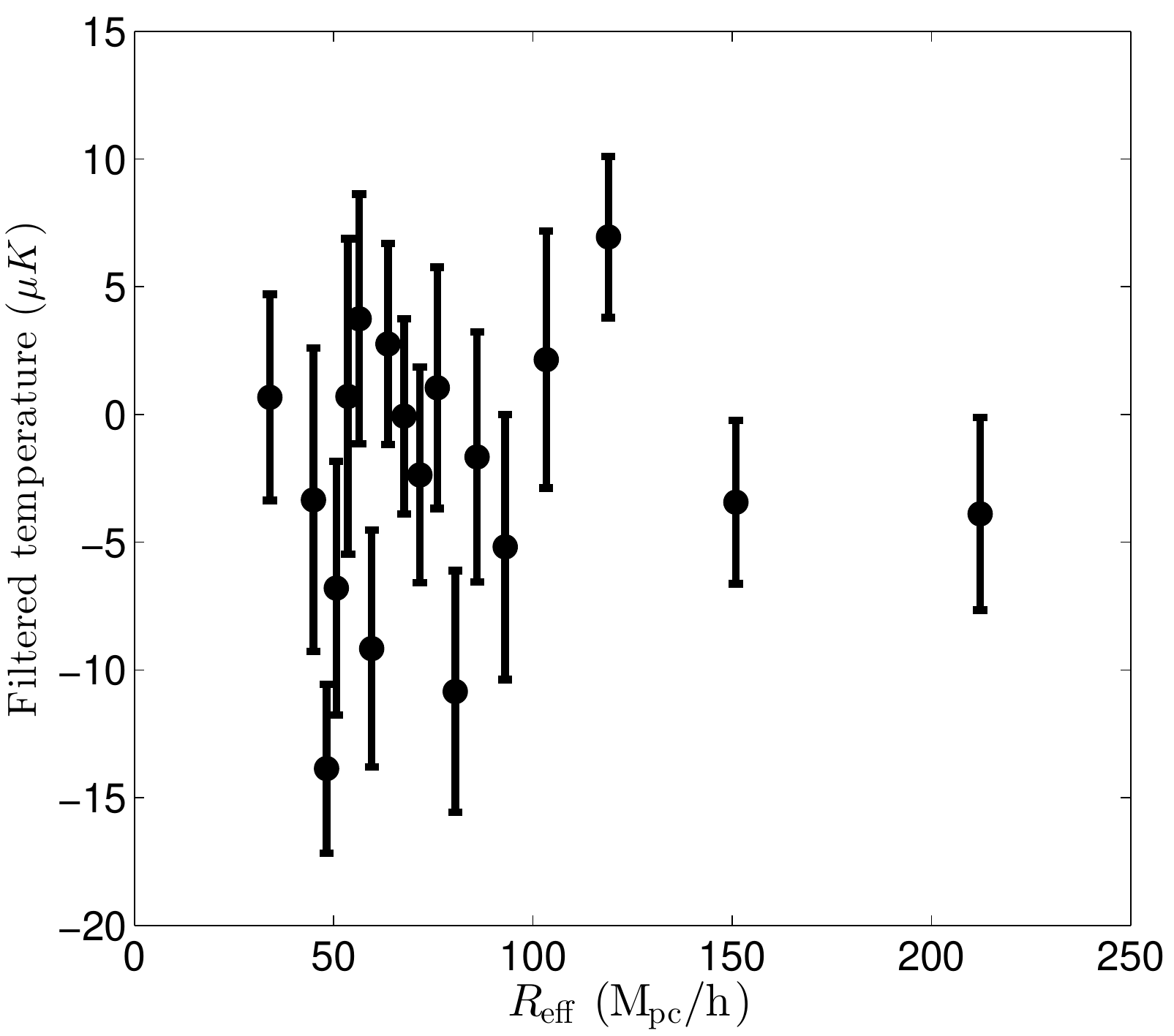} 
  \caption{The average filtered ISW temperature shift caused by voids identified in the SDSS data by \citet{Cai:2013ik} (C13), as a function of density ratio parameter (top) and effective radius (bottom).}
  \label{fig:Caiscatter}
\end{figure}

Secondly, we observe that the C13 result \emph{cannot} be seen as a confirmation, or a reproduction, of the higher significance G08 result. This is because the methodology of the two studies is rather different: G08 used a catalogue of superstructures selected on the basis of density ratio $r$ alone, whereas C13 select their catalogue on the basis of $R_{\rm eff}$ and not $r$. Imposing $r\geq2$ (the G08 selection criterion for voids) on the C13 catalogue reduces the observed S/N (as we see below). Therefore the correct interpretation of the results in C13 are that they are a failure to reproduce the signal seen in G08 and a tentative detection of some other new signal (with low significance). 

C13 also observe that the filtered temperature signal from their sample of voids decreases (becomes more negative) with increasing values of $r$ and $R_\rmn{eff}$. This is shown in Figure 5 of that paper. While cautioning against over-interpretation of this trend, they view it as an encouraging sign that the observed signal may be due to physical effects rather than a random statistical fluctuation (this argument is the same as that in Section~\ref{sec:ISWsimsum}). However, we believe that the interpretation of this figure is not so straightforward.

The first reason for this is that the plot of temperatures shown in Figure 5 of C13 uses \emph{cumulative} bins, which means that individual pixels are highly correlated with each other. Therefore the existence of an apparent trend in pixel temperatures, decreasing towards the upper right of the plot, follows simply from the existence of a single cold pixel in the upper right corner. The probability that this pixel should be cold simply due to a chance random fluctuation is relatively large (the S/N for this choice of $r$ and $R_\rmn{eff}$ cuts is not in itself significant). Given that this particular pixel is cold, the probability of seeing an apparent trend in temperature towards this pixel in such a cumulative plot, even in the absence of any real physical effect, is close to unity.

It is also true that the appearance of this apparent gradient in stacked temperature with $r$ and $R_\rmn{eff}$ is sensitive to the extent of the plot. In Figure~\ref{fig:Caiheat} we demonstrate this by reproducing Figure 5 of C13 but extending the borders in both directions (the original figure extended only as far as the ratio value 2 on both axes). It is clear that the apparent gradient of stacked temperature does not continue: in fact for the largest radii and density ratios the stack of C13 voids produces a net \emph{positive} temperature shift. A similar effect is seen if Figure 6 of C13 is extended to larger radii.

Finally, Figure~\ref{fig:Caiscatter} shows scatter plots of $\overline{\Delta T}$ as a function of both $r$ and $R_{\rm eff}$ for the (cut) C13 void catalogue, similar to Figures~\ref{fig:effrad} and \ref{fig:densrat}. There is no apparent trend in the temperature signal with either property alone, arguing against the observed effect having a physical origin.

Of course, none of these observations should take away from the fact that, using the \emph{a priori} cuts chosen by calibration to simulation, C13 saw a stacked temperature signal with the correct sign to be interpreted as an ISW effect and a statistical significance of $\sim2.3\sigma$. The caveats we have added here are to do with the physical interpretation of this (low-significance) detection, but do not change the fact of its existence. Nor can our own null detection conclusively exclude the possibility that the original G08 observation has some physical significance rather than simply being an unlikely fluke. Only time and observations of other volumes of space will be able to resolve this question.

\section{Summary and Conclusions}
\label{sec:conclude}

In this work we have examined the stacked integrated Sachs-Wolfe temperature signal in the cosmic microwave background due to cosmic superstructures (voids and superclusters). 

Using simulated ISW maps and mock LRG catalogues from the Jubilee ISW Project $N$-body simulation, we first examined the \Lam CDM expectation for the stacked signal from voids and superclusters in the galaxy distribution identified using the structure-finding algorithm {\small ZOBOV}. Jubilee consists of $6000^3$ particles in a volume of $(6 h^{-1} {\rm Gpc})^3$, with a minimum resolved halo-mass of $\simeq 1.5 \times 10^{12} h^{-1} M_\odot$. This resolution allows the creation of realistic \emph{full-sky} mock catalogues of luminous red galaxies, which meant that we were able, for the first time to match the simulation methodology to that used in observations. At the same time, because of the large box size of the simulation, which is complete over the whole sky out to a redshift of $z=1.4$, we can capture the complete ISW signal without needing to tile the box. We can therefore be confident that our simulation is not missing any power on the largest scales.

Our study of the Jubilee data confirms that superstructures do contribute an ISW temperature shift, but that the \Lam CDM expectation for its amplitude is extremely small and the signal should therefore be unmeasurable in the CMB. This confirms earlier theoretical estimates \citep{Hunt:2008wp,Nadathur:2011iu,Flender:2012wu} and results from smaller simulations \citep{HernandezMonteagudo:2012ms,Cai:2013ik}. This means that the high-significance ($>4\sigma$) detection of such a signal reported by \cite{Granett:2008ju}, if due to a true physical effect, is a sign of some unknown new physics beyond the \Lam CDM model.

In order to determine whether this signal is due to a physical effect, we searched for evidence of a similar temperature effect in stacked images of the Planck CMB along directions of superstructures identified in SDSS DR7 spectroscopic galaxy surveys. Unlike some previous studies, we use a robustly identified catalogue of genuine voids for this task; also in contrast to recent detection attempts (but in keeping with the original measurement by \citealt{Granett:2008ju}), we include supercluster directions in our search in order to increase the possibility of detection.

Our results show a signal amplitude consistent with zero, i.e. a null detection, when using the full superstructure catalogues. Applying several different physically-motivated cuts to the catalogues does not increase the signal, nor do we see anything when exactly reproducing the superstructure selection criteria employed by \citet{Granett:2008ju}. We conclude that an analogous effect to that seen in the original observation does not exist in this independent data set. 

To reconcile our results with a physical interpretation of the original high-significance claim of a detection, one would require the hypothetical new physics that explains that result to be either strongly redshift-dependent, or dependent on some other peculiar property of the photometric galaxy sample used for that detection. We stress that our own null detection is perfectly in keeping with the \Lam CDM expectation of an undetectable signal.

Finally, we briefly discussed our result in light of another recent tentative claim \citep{Cai:2013ik} of a detectable temperature shift (albeit at low significance, $\sim2\sigma$) found using a slightly different catalogue of voids drawn from the same data. This observation, if due to physical effects, would also be in conflict with \Lam CDM. Although we reproduce this result when using this catalogue of voids, we believe that it cannot be claimed to support the original \citet{Granett:2008ju} result because the methodology used in the two studies differs significantly, and when the original methodology is applied to the new data, even the tentative hints of a signal disappear. We also argue that the temperature effect seen by \cite{Cai:2013ik} does not show the same behaviour with changes in the void parameters as would be expected for a physical effect.

Nonetheless, the initial high-significance result still remains unexplained. We have shown that it is certainly not due to an integrated Sachs-Wolfe effect in \Lam CDM and does not appear to exist at a lower redshift. However, a conclusive determination of whether it is the sign of interesting new physics, or simply a very rare statistical fluke will require the use of new galaxy survey data at redshifts $z\gtrsim0.5$ in sky regions complementary to SDSS.

\section{Acknowledgements}

We gratefully acknowledge the contributions of R. B. Barreiro, J. M. Diego, J. Gonz\'alez-Nuevo, A. Marcos-Caballero, E. Mart\'inez-Gonz\'alez and P. Vielva to the Jubilee ISW project. The Jubilee simulation was performed on the Juropa supercomputer of the J\"ulich Supercomputing Centre (JSC).

SH and ITI were supported by the Science and Technologies Facilities Council (grant number ST/I000976/1). The research leading to these results has received funding from the European Research Council under the European Union’s Seventh Framework Programme (FP/2007–2013) / ERC Grant Agreement No. [308082]. SN acknowledges support from Academy of Finland grant 1263714. AK is supported by the {\it Ministerio de Econom\'ia y Competitividad} (MINECO) in Spain through grant AYA2012-31101 as well as the Consolider-Ingenio 2010 Programme of the {\it Spanish Ministerio de Ciencia e Innovaci\'on} (MICINN) under grant MultiDark CSD2009-00064. He also acknowledges support from the {\it Australian Research Council} (ARC) grants DP130100117 and DP140100198. GY acknowledges support from MINECO (Spain) under research grants AYA2012-31101 and FPA2012-34694.

This research has used data from the SDSS Data Release 7. Funding for the SDSS and SDSS-II has been provided by the Alfred P. Sloan Foundation, the Participating Institutions, the National Science Foundation, the U.S. Department of Energy, the National Aeronautics and Space Administration, the Japanese Monbukagakusho, the Max Planck Society, and the Higher Education Funding Council for England. The SDSS website is \url{http://www.sdss.org/}. Some of the results in this paper have been derived using the HEALPix package.

\bibliography{../refs}
\bibliographystyle{mn2e}

\label{lastpage}

\end{document}